\documentclass[aps,prx,twocolumn,notitlepage,superscriptaddress]{revtex4-2}
\usepackage[ruled, linesnumbered]{algorithm2e}
\usepackage{tikz}
\usetikzlibrary{arrows.meta, positioning}
\usepackage{bm} 
\usepackage{graphicx}
\usepackage{dcolumn}
\newcolumntype{d}[1]{D{.}{.}{#1}}
\usepackage{amsmath,amsthm,stmaryrd,amssymb,array} 
\usepackage{color}
\usepackage{hyperref}
\usepackage{booktabs}
\usepackage{dsfont}
\usepackage{tikz}
\usetikzlibrary{calc}
\usepackage{siunitx}
\hypersetup{colorlinks,
            citecolor=red,
            urlcolor=blue,
            bookmarks=false,
            hypertexnames=true}

\DeclareMathOperator*{\Tr}{Tr}
\DeclareMathOperator*{\Var}{Var}

\newcommand{\ket}[1]{\vert{#1}\rangle}

\newcommand{\braket}[1]{\langle{#1}\rangle}

\definecolor{LB}{RGB}{134,41,198}

\usepackage{soul}

\definecolor{bluelink} {RGB}{24, 95, 165}
\definecolor{redlink}  {RGB}{180, 40, 30}
\definecolor{nodecolor}{RGB}{100,100,100}
\definecolor{dashcolor} {RGB}{140,140,140}
 
\tikzset{
  site/.style={
    circle, draw=nodecolor, fill=white,
    line width=1pt,
    minimum size=24pt, inner sep=0pt,
    font=\normalsize
  },
  neglink/.style={
    -{Stealth[length=7pt, width=5pt]},
    line width=1.5pt, color=redlink
  },
  dashlink/.style={
    -{Stealth[length=7pt, width=5pt, open]},
    line width=1pt,
    color=dashcolor,
    dash pattern=on 5pt off 3pt
  },
}

\begin{document}

\title{Hardware-Aware Fermion-to-Qubit Mappings for Simulating the 2D Hubbard Model on Heavy-Hexagon Quantum Processors}

\author{Alessio Esposito} \email{a.esposito82@campus.unimib.it}
\affiliation{Department of Physics, University of Milano-Bicocca,
Piazza della Scienza 3, I-20126 Milano-Bicocca (MI), Italy}

\author{Andrea Giachero} \email{andrea.giachero@unimib.it}
\affiliation{Department of Physics, University of Milano-Bicocca,
Piazza della Scienza 3, I-20126 Milano-Bicocca (MI), Italy}
\affiliation{INFN Milano-Bicocca,
Piazza della Scienza 3, I-20126 Milano-Bicocca (MI), Italy}
\affiliation{Bicocca Quantum Technologies (BiQuTe) Centre, Piazza della Scienza 3, I-20126 Milano-Bicocca (MI), Italy}

\author{Zolt\'an Zimbor\'as}\email{zoltan.zimboras@helsinki.fi}
\affiliation{University of Helsinki, Yliopistonkatu 4 00100 Helsinki, Finland}
\affiliation{HUN-REN Wigner Research Centre for Physics, Budapest, Hungary}
\affiliation{Algorithmiq Ltd, Kanavakatu 3C 00160 Helsinki, Finland}

\author{Leonardo Banchi } \email{leonardo.banchi@unifi.it}
\affiliation{Department of Physics and Astronomy, University of Florence,
via G. Sansone 1, I-50019 Sesto Fiorentino (FI), Italy}
\affiliation{ INFN Firenze, via G. Sansone 1, I-50019, Sesto Fiorentino (FI), Italy}

\begin{abstract}
Quantum simulation of strongly correlated fermionic systems is among the most promising near-term applications of quantum computing, but its practical efficiency depends critically on the choice of fermion-to-qubit mapping and on the connectivity of the underlying hardware. In this work we address this problem in the context of the two-dimensional Hubbard model, simulated on IBM superconducting quantum processors with heavy-hexagon connectivity. We numerically benchmark the Jordan–Wigner, Bravyi–Kitaev, and Bonsai transformations, evaluating their Pauli weight and SWAP overhead across seven heavy-hexagon chips of increasing size. We show that, while the Bravyi–Kitaev mapping initially exhibits a lower Pauli weight, this advantage is eliminated once routing costs are taken into account, confirming the Bonsai mapping as the most hardware-efficient baseline transformation for this architecture. We then use the Bonsai mapping to construct the qubit Hamiltonian of the 2D spinful Fermi-Hubbard model, introducing a simulated annealing algorithm that optimizes the assignment of Majorana strings to lattice sites, reducing the cost function by nearly 50\%. Finally, we simulate on quantum hardware the time evolution of fermionic states  up to $6\times6 $ lattices, confirming the viability of the Bonsai encoding for hardware-aware large-scale two-dimensional simulations. 
\end{abstract}

\maketitle

\section{INTRODUCTION}
    
Since Feynman's idea of simulating quantum systems using devices that operate according to the laws of quantum mechanics \cite{Feynman},
the field of quantum computing has experienced remarkable progress, leading to the realization of quantum processors with hundreds and, more recently, thousands of qubits \cite{AtomComputing_2023,IBM_Condor_2023}. Among the most promising applications of quantum computation is the simulation of condensed matter systems, particularly those involving interacting many-fermion models \cite{georgescu2014quantum}.

The feasibility of simulating fermionic systems on quantum computers arises from the possibility of representing fermionic degrees of freedom using qubits through suitable fermion-to-qubit mappings. Such mappings allow one to encode arbitrary superpositions of electronic configurations within the computational basis of a quantum register. Once a suitable reference state—such as a Slater determinant corresponding to the fermionic ground state—is prepared, fermionic creation and annihilation operators can be expressed in terms of Pauli operators acting on qubits. This construction enables the mapping of fermionic Fock states onto computational basis states, while entanglement is not required for the preparation of uncorrelated reference states \cite{Miller}.

Despite the formal equivalence between fermionic and qubit Hilbert spaces, the practical implementation of fermionic simulations critically depends on the choice of the fermion-to-qubit transformation. Any valid mapping must reproduce the fermionic anticommutation relations, while at the same time minimizing the complexity of the resulting qubit operators. In particular, fermionic operators are mapped to strings of Pauli operators whose length—commonly referred to as the \textit{Pauli weight}—directly affects the depth of quantum circuits and their susceptibility to noise and decoherence. Reducing the Pauli weight of these strings is therefore a key objective in the design of efficient fermionic simulations.

Over the past years, several fermion-to-qubit mappings have been proposed with the aim of reducing the Pauli weight of the encoded operators. Among the most widely used transformations are the Jordan–Wigner (JW) \cite{JW}, Bravyi–Kitaev (BK) \cite{Bravyi}, and parity mappings \cite{bravyi2017taperingqubitssimulatefermionic}, which are now implemented in major quantum software libraries. More recently, non-linear encodings have been introduced to further reduce the number of qubits required to represent a given fermionic system by exploiting symmetries of the underlying Hamiltonian, leading to mappings where the number of qubits involved can be smaller than the number of fermionic modes ($n<m$) \cite{bravyi2017taperingqubitssimulatefermionic,Steudtner_2018,fischer2019symmetryconfigurationmappingrepresenting}. While these approaches can significantly reduce the required physical resources, their effectiveness strongly depends on the structure of the system under consideration.

Even when considering mappings with favorable asymptotic scaling—such as the $\mathcal{O}(n)$ scaling of the Jordan–Wigner transformation or the $\mathcal{O}(\log_2 n)$ scaling of the Bravyi–Kitaev mapping—their practical efficiency can be severely affected by the topology of the physical quantum hardware. Current quantum processors, including IBM’s superconducting devices, support operations only through a finite set of native single- and two-qubit gates. Crucially, two-qubit gates can be applied only between adjacent qubits in the hardware connectivity graph. As a consequence, Pauli strings acting on non-contiguous sets of qubits must be decomposed into sequences of native gates involving additional \text{\footnotesize SWAP} operations to bring qubits next to each other. These \text{\footnotesize SWAP} gates introduce a significant overhead in circuit depth and noise, potentially nullifying the theoretical advantages of a given fermion-to-qubit mapping \cite{Miller}.

Motivated by this issue, several works have investigated strategies to incorporate hardware connectivity into the design of fermion-to-qubit mappings. In particular, an efficient mapping tailored to specific hardware topologies was introduced in \cite{Miller} through the Bonsai algorithm. This method employs a greedy procedure that starts from a chosen root qubit and iteratively connects at most three neighboring qubits, referred to as descendant qubits, thereby constructing a tree graph that spans the entire set of qubits available on the chip. The algorithm is especially well suited to hardware layouts such as the heavy-hexagon topology adopted by IBM superconducting quantum processors \cite{IBM_HeavyHex_2021}.

The resulting structure corresponds to a ternary tree mapping derived from the balanced ternary tree construction proposed in \cite{Jiang}. While the ideal scaling of the Pauli strings associated with ternary tree mappings is $\mathcal{O}(\log_3 n)$, topological constraints imposed by the heavy-hexagon lattice lead to an effective scaling of $\mathcal{O}(\sqrt{n})$ in this specific case \cite{Miller}.

In the first part of this work, we numerically benchmark three fermion-to-qubit mappings—the Jordan–Wigner, Bravyi–Kitaev, and Bonsai transformations—focusing on their performance under the heavy-hexagon connectivity. We analyze both quadratic and quartic fermionic terms appearing in the Hubbard Hamiltonian and evaluate the corresponding Pauli weights and \text{\footnotesize SWAP} overheads for seven heavy-hexagon-based chips with increasing numbers of qubits. From these data, we extract the asymptotic behavior of each mapping by fitting the measured Pauli weights and \text{\footnotesize SWAP} counts associated with single and double fermionic excitations.

Our results show that, while the Bravyi–Kitaev mapping initially exhibits a lower Pauli weight compared to the Bonsai transformation, this advantage is lost once the \text{\footnotesize SWAP} overhead induced by the hardware connectivity is taken into account. Both Jordan–Wigner and Bravyi–Kitaev mappings display an overall linear scaling in the combined cost, whereas the Bonsai mapping requires no \text{\footnotesize SWAP} operations for single excitations and exhibits a cubically reduced number of \text{\footnotesize SWAP}s for quartic terms relative to the other mappings, in agreement with the predictions of \cite{Miller}. These findings confirm that, for heavy-hexagon architectures, the Bonsai mapping provides the most efficient baseline transformation among the mappings considered.

Building on these results, we employ the Bonsai mapping to construct the qubit Hamiltonian of the two-dimensional Hubbard model on a square lattice, considering spinful fermions. In this context, further optimizations become possible, as different assignments of Pauli strings to lattice sites can lead to cancelations in the single- and double-excitation terms of the fermionic Hamiltonian. A simulated annealing algorithm was implemented to further optimize the Bonsai mapping. The algorithm explores a limited search space consisting of swaps of Pauli string assignments among fermionic sites, resulting in a reduced cost function of nearly $50\%$ for both spinless and spinful cases.

Based on the previous results, we can further address the following question: \textit{is the Bonsai Hamiltonian the most efficient fermion-to-qubit representation when hardware connectivity constraints are explicitly taken into account?} In the conclusions, we suggest to extend previous Clifford-based heuristic optimization approaches \cite{Yu_2025} by introducing a cost function that explicitly includes the \text{\footnotesize SWAP} overhead induced by the heavy-hexagon connectivity of superconducting quantum processors and the errors associated to every single qubit on the hardware. By navigating the space of fermion-to-qubit mappings through Clifford transformations under this hardware-aware cost function, it is possible to investigate whether alternative mappings can outperform the Bonsai transformation in terms of total circuit cost and robustness to noise.

Finally, as discussed in Sec. \ref{Sec: simulation}, we employ the improved qDRIFT \cite{Campbell_2019} protocol proposed in \cite{david2025tightererrorboundsqdrift} to simulate the time evolution of fermionic states and extract physical observables of the two-dimensional Hubbard model. We consider square lattices comprising 32 and 72 spinful fermions. For both cases, we compute on-site and global occupations, magnetization, energy, spatial spin–spin correlation functions, and structure factors. We conclude with an outlook on possible extensions of this approach and its implications for the quantum simulation of strongly and weakly correlated condensed matter systems.

\section{Notation and State of the Art}

In this section, we summarize the main concepts and results relevant to the simulation of many-fermion systems on quantum computers, which will be used throughout this work.

Let us consider a system of $n$ fermionic modes. Fermionic creation and annihilation operators satisfy the canonical anticommutation relations
\begin{equation}
    \{a_i,a_j\}=\{a^{\dagger}_i,a^{\dagger}_j\}=0, 
    \qquad 
    \{a_i,a^{\dagger}_j\}=\delta_{ij}\mathds{1},
    \label{anticommute}
\end{equation}
where $i,j=0,\dots,n-1$.

It is often convenient to reformulate these relations in terms of Majorana operators, defined as
\begin{equation}
m_{2j}=a^{\dagger}_j+a_j,
\qquad 
m_{2j+1}=i(a^{\dagger}_j-a_j),
\end{equation}
where $j=0,\dots,n-1$.
These operators satisfy
\begin{equation}
    \{m_i,m_j\}=2\delta_{ij}\mathds{1}.
\end{equation}
Majorana operators are Hermitian, square to the identity, and are algebraically independent, meaning that no Majorana operator can be expressed as a polynomial function of the remaining ones.

An encoding of $n$ fermionic modes into $m$ qubits is defined as a unitary embedding
\begin{equation}
    J:\mathcal{H}_f \rightarrow \mathcal{B}^{\otimes m},
\end{equation}
where $\mathcal{H}_f$ and $\mathcal{B}^{\otimes m}$ denote the fermionic and $m$-qubits Hilbert spaces, respectively. Under this mapping, a fermionic state $\ket{\psi}_f$ is mapped to a qubit state $\ket{\phi}_q = J\ket{\psi}_f$. A fermionic unitary operator $U_f \in \boldsymbol{L}(\mathcal{H}_f)$ is represented on the qubit space by a unitary $U_q \in \boldsymbol{L}(\mathcal{B}^{\otimes m})$ if
\begin{equation}
    U_q = J U_f J^{-1},
\end{equation}
which guarantees that the fermionic anticommutation relations in Eq.~(\ref{anticommute}) are preserved.

Throughout this work, we restrict our attention to linear fermion-to-qubit mappings and therefore set $n=m$. Under such mappings, fermionic operators can be expressed as linear combinations of \emph{Pauli strings}, defined as tensor products of single-qubit Pauli operators $P_i \in \{ \mathds{1}, X, Y, Z \}$ acting on the $i$-th qubit, where $i=0,\dots,n-1$. A key figure of merit for the efficiency of a mapping is the \emph{Pauli weight} of a string $S$, defined as the number of qubits on which it acts nontrivially:
\begin{equation}
    \omega(S)= \sum_{i=0}^{n-1} \alpha_i,
    \qquad 
    \alpha_i =
    \begin{cases}
        0 & \text{if } P_i=\mathds{1},\\
        1 & \text{otherwise}.
    \end{cases}
    \label{eq:pauli_weight}
\end{equation}

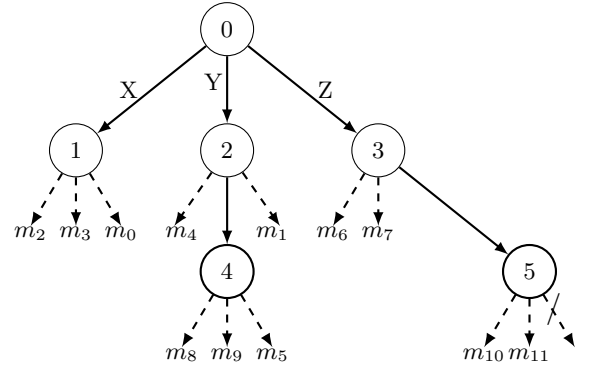
\begin{figure}
    \centering
    \begin{tikzpicture}[
        level distance=1.6cm,
        sibling distance=2cm,
        every node/.style={draw,circle,minimum size=7mm,inner sep=0pt,font=\small},
        edge from parent/.style={draw,-latex,thick},
        leg/.style={draw,-latex,dashed,thick}
    ]
    
    \node (r) {0}
      child {
        node (a) {1}    
        child[missing]
        child[missing]
        child[missing]
      }
      child {
        node (b) {2}     
        child[missing]
        child {
          node (d) {4}
          edge from parent
          child[missing]
          child[missing]
          child[missing]
        }
        child[missing]
      }
      child {
        node (c) {3}    
        child[missing]
        child[missing]
        child {
          node (e) {5}
          edge from parent
          child[missing]
          child[missing]
          child[missing]
        }
      };
    
    \draw[leg] (a) -- ++(-0.6,-1.0);
    \draw[leg] (a) -- ++(0.0,-1.0);
    \draw[leg] (a) -- ++(0.6,-1.0);
    
    \draw[leg] (b) -- ++(-0.7,-1.0);
    \draw[leg] (b) -- ++(0.7,-1.0);
    
    \draw[leg] (c) -- ++(-0.6,-1.0);
    \draw[leg] (c) -- ++(0.0,-1.0);
    
    \draw[leg] (d) -- ++(-0.6,-1.0);
    \draw[leg] (d) -- ++(0.0,-1.0);
    \draw[leg] (d) -- ++(0.6,-1.0);
    
    \draw[leg] (e) -- ++(-0.6,-1.0);
    \draw[leg] (e) -- ++(0.0,-1.0);
    \draw[leg] (e) -- ++(0.6,-1.0);
    \node[font=\large, draw=none] at ($(e)+(0.32,-0.5)$) {/};
    
    \node[font=\small, draw=none] at ($(r)!0.5!(a)+(-0.3,0)$) {X};
    \node[font=\small, draw=none] at ($(r)!0.5!(b)+(-0.18,0.1)$) {Y};
    \node[font=\small, draw=none] at ($(r)!0.5!(c)+(0.3,0)$) {Z};
    
    \node[font=\small, draw=none] at ($(a)+(-0.6,-1.1)$) {$m_2$};
    \node[font=\small, draw=none] at ($(a)+(0.0,-1.1)$) {$m_3$};
    \node[font=\small, draw=none] at ($(a)+(0.6,-1.1)$) {$m_0$};
    
    \node[font=\small, draw=none] at ($(b)+(-0.6,-1.1)$) {$m_4$};
    \node[font=\small, draw=none] at ($(b)+(0.6,-1.1)$) {$m_1$};
    
    \node[font=\small, draw=none] at ($(d)+(-0.6,-1.1)$) {$m_8$};
    \node[font=\small, draw=none] at ($(d)+(0.0,-1.1)$) {$m_9$};
    \node[font=\small, draw=none] at ($(d)+(0.6,-1.1)$) {$m_5$};
    
    \node[font=\small, draw=none] at ($(c)+(-0.6,-1.1)$) {$m_6$};
    \node[font=\small, draw=none] at ($(c)+(0,-1.1)$) {$m_7$};
    
    \node[font=\small, draw=none] at ($(e)+(-0.6,-1.1)$) {$m_{10}$};
    \node[font=\small, draw=none] at ($(e)+(0,-1.1)$) {$m_{11}$};
    
    \end{tikzpicture}
    \caption{A TT example consisting of 6 qubits. Every node must have three links (edges or legs). The outgoing links from a node are associated with the $X,Y,Z$ label. Every Majorana string stems from one of the paths from the root to a leg, and to guarantee the algebraic independence, the path composed of $Z$ links is ruled out. The remaining strings are coupled together as discussed.}
    \label{fig:TT example}
\end{figure}

\subsection{Ternary tree structure}

We now briefly review the ternary tree (TT) structure, which provides a convenient graphical representation for several fermion-to-qubit mappings. A TT is a graph $\mathcal{G}=(V,L)$ composed of a set of nodes $V$ connected by a set of links $L$. The tree is constructed starting from a root node, to which at most three descendant nodes are attached. This procedure is iterated recursively to generate successive levels of the tree.

The resulting graph satisfies the following properties:
\begin{enumerate}
    \item Each node (except the root) is connected to a unique ancestor node at the previous level.
    \item Each node has at most three downward links; missing links are completed by \emph{legs}, which do not connect to further nodes.
    \item The graph is loop-free, ensuring that no node is connected more than once.
\end{enumerate}

A \emph{path} $\mathbf{p}$ in the tree is defined as an ordered sequence of downward links
$\mathbf{p}=\{L_0,\dots,L_{l-1}\}$, where each link $L_k$ connects node $C_k$ to node $C_{k+1}$
for $k=0,\dots,l-2$, while the final link $L_{l-1}$ is a \emph{leg} attached to node $C_{l-1}$.
The associated sequence of nodes is therefore
$\mathbf{c}=\{C_0,\dots,C_{l-1}\}$.
In a ternary tree with $n$ nodes, there are $2n+1$ legs and, consequently, $2n+1$ distinct paths connecting the root to a leg.

Without loss of generality, nodes are enumerated from left to right starting from the root, and each node is associated with a qubit. The three downward links of each node are labeled with the Pauli operators $X$, $Y$, and $Z$. This labeling establishes a direct correspondence between paths in the tree and Pauli strings acting on the qubit register, as illustrated in Fig. \ref{fig:TT example}. Within this framework, Majorana operators can be mapped to Pauli strings derived from these paths \cite{Jiang,Vlasov_2022}.

Although a TT with $n$ nodes generates $2n+1$ Pauli strings, only $2n$ Majorana operators are required. The resulting set of Pauli strings is therefore algebraically dependent; however, it can be shown that removing any one string yields a set of $2n$ independent operators \cite{Miller}. In practice, the string corresponding to the path labeled exclusively by $Z$ operators is commonly discarded.

To associate Majorana operators $m_{2i}$ and $m_{2i+1}$ with Pauli strings (where the $i$ refers to the associated $i$-th qubit or node), we follow the procedure introduced in \cite{Jiang,Vlasov_2022}. For a given node, we first follow its downward $X$ link; if the link is not a leg, we continue descending along $Z$-labeled links until a leg is reached. Repeating the procedure starting from the $Y$ link produces a second Pauli string. These two strings can be chosen to represent the corresponding pair of Majorana operators, as explained in Fig. \ref{fig:TT example}.

\begin{figure*}[t]
    \centering

    
    \begin{minipage}[t]{0.45\textwidth}
        \vspace{0pt}
        \centering
        \scalebox{0.75}{%
        \begin{tikzpicture}[
            level distance=1.6cm,
            sibling distance=2cm,
            every node/.style={draw,circle,minimum size=7mm,inner sep=0pt,font=\small},
            edge from parent/.style={draw,-latex,thick},
            leg/.style={draw,-latex,dashed,thick}
        ]
        \node (r) {0}
          child[missing]
          child[missing]
          child {
            node (a) {1}    
            child[missing]
            child[missing]
            child {
            node (b) {2}     
            child[missing]
            child {
              node (c) {3}
              edge from parent
              child[missing]
              child[missing]
              child[missing]
            }
           }    
          };
        
        \draw[leg] (r) -- ++(-0.6,-1.0);
        \draw[leg] (r) -- ++(0.0,-1.0);
        
        \draw[leg] (a) -- ++(-0.6,-1.0);
        \draw[leg] (a) -- ++(0.0,-1.0);
       
        \draw[leg] (b) -- ++(-0.7,-1.0);
        \draw[leg] (b) -- ++(0,-1.0);
        
        \draw[leg] (c) -- ++(-0.6,-1.0);
        \draw[leg] (c) -- ++(0.0,-1.0);
        \draw[leg] (c) -- ++(0.6,-1.0);

        \node[font=\large, draw=none] at ($(c)+(0.32,-0.5)$) {/};
        
        \node[font=\small, draw=none] at ($(r)+(-0.6,-0.5)$) {X};
        \node[font=\small, draw=none] at ($(r)+(-0.15,-0.7)$) {Y};
        \node[font=\small, draw=none] at ($(r)!0.5!(a)+(-0.35,0.05)$) {Z};

        \node[font=\small, draw=none] at ($(r)+(-0.6,-1.1)$) {$m_0$};
        \node[font=\small, draw=none] at ($(r)+(0,-1.1)$) {$m_1$};
        
        \node[font=\small, draw=none] at ($(a)+(-0.6,-1.1)$) {$m_2$};
        \node[font=\small, draw=none] at ($(a)+(0.0,-1.1)$) {$m_3$};
         
        \node[font=\small, draw=none] at ($(b)+(-0.6,-1.1)$) {$m_4$};
        \node[font=\small, draw=none] at ($(b)+(0,-1.1)$) {$m_5$};
        
        \node[font=\small, draw=none] at ($(c)+(-0.6,-1.1)$) {$m_6$};
        \node[font=\small, draw=none] at ($(c)+(0.0,-1.1)$) {$m_7$};

        \end{tikzpicture}}
    \end{minipage}%
    \hfill 
    \begin{minipage}[t]{0.45\textwidth}    
        \vspace{0pt}
        \centering
        \scalebox{0.75}{
        \begin{tikzpicture}[
            level distance=1.6cm,
            sibling distance=1.5cm,
            every node/.style={draw,circle,minimum size=7mm,inner sep=0pt,font=\small},
            edge from parent/.style={draw,-latex,thick},
            leg/.style={draw,-latex,dashed,thick}
        ]
        \node (r) {0}
          child {
            node (a) {1}    
            child [sibling distance=2cm]{
            node (b) {2}     
            child [sibling distance=1.0cm]{
              node (d) {4}
              edge from parent
              child[missing]
              child[missing]
              child[missing]
             }
             child[missing]
             child [sibling distance=1.0cm]{
              node (f) {5}
              edge from parent
              child[missing]
              child[missing]
              child[missing]
             }
            }
            child[missing]
            child [sibling distance=2cm]{
            node (c) {3}    
            child [sibling distance=1.0cm]{
              node (g) {6}
              edge from parent
              child[missing]
              child[missing]
              child[missing]
             }
            child[missing]
            child [sibling distance=1.0cm]{
              node (e) {7}
              edge from parent
              child[missing]
              child[missing]
              child[missing]
            }
           }
          }
          child[missing]
          child[missing];
        
        \draw[leg] (r) -- ++(0.6,-1.0);
        \draw[leg] (r) -- ++(0.0,-1.0);
        
        \draw[leg] (a) -- ++(0.0,-1.0);
         
        \draw[leg] (b) -- ++(0.0,-1.0);
        
        \draw[leg] (c) -- ++(0.0,-1.0);
        
        \draw[leg] (d) -- ++(-0.6,-1.0);
        \draw[leg] (d) -- ++(0.0,-1.0);
        \draw[leg] (d) -- ++(0.6,-1.0);
        
        \draw[leg] (e) -- ++(-0.6,-1.0);
        \draw[leg] (e) -- ++(0.0,-1.0);
        \draw[leg] (e) -- ++(0.6,-1.0);

        \draw[leg] (f) -- ++(-0.6,-1.0);
        \draw[leg] (f) -- ++(0.0,-1.0);
        \draw[leg] (f) -- ++(0.6,-1.0);

        \draw[leg] (g) -- ++(-0.6,-1.0);
        \draw[leg] (g) -- ++(0.0,-1.0);
        \draw[leg] (g) -- ++(0.6,-1.0);
           
        \node[font=\small, draw=none] at ($(r)!0.33!(a)+(-0.3,0)$) {X};
        \node[font=\small, draw=none] at ($(r)+(-0.18,-0.6)$) {Y};
        \node[font=\small, draw=none] at ($(r)+(0.4,-0.6)$) {/};

        \node[font=\small, draw=none] at ($(r)+(0,-1.1)$) {$m_1$};
        
        \node[font=\small, draw=none] at ($(a)+(0,-1.1)$) {$m_3$};
        
        \node[font=\small, draw=none] at ($(b)+(0,-1.1)$) {$m_4$};
       
        \node[font=\small, draw=none] at ($(d)+(-0.6,-1.1)$) {$m_8$};
        \node[font=\small, draw=none] at ($(d)+(0.0,-1.1)$) {$m_9$};
        \node[font=\small, draw=none] at ($(d)+(0.6,-1.1)$) {$m_5$};
        
        \node[font=\small, draw=none] at ($(c)+(0,-1.1)$) {$m_6$};
        
        \node[font=\small, draw=none] at ($(e)+(-0.6,-1.1)$) {$m_{14}$};
        \node[font=\small, draw=none] at ($(e)+(0,-1.1)$) {$m_{15}$};
        \node[font=\small, draw=none] at ($(e)+(0.6,-1.1)$) {$m_0$};
        
        \node[font=\small, draw=none] at ($(f)+(-0.6,-1.1)$) {$m_{10}$};
        \node[font=\small, draw=none] at ($(f)+(0.0,-1.1)$) {$m_{11}$};
        \node[font=\small, draw=none] at ($(f)+(0.6,-1.1)$) {$m_2$};

        \node[font=\small, draw=none] at ($(g)+(-0.6,-1.1)$) {$m_{12}$};
        \node[font=\small, draw=none] at ($(g)+(0.0,-1.1)$) {$m_{13}$};
        \node[font=\small, draw=none] at ($(g)+(0.6,-1.1)$) {$m_7$};
        \end{tikzpicture}
        }
    \end{minipage}%

    \caption{Two ternary tree (TT) examples. The left panel shows a TT with 4 qubits that maps to the JW mapping, the right panel a TT with 8 qubits that maps to the BK mapping. The first TT has a number of levels of the tree equal to the number of qubits, then the corresponding Pauli strings are linear with the scalability, whereas the second one is logarithmic with the number of qubits.}
    \label{fig:TT_examples}
\end{figure*}
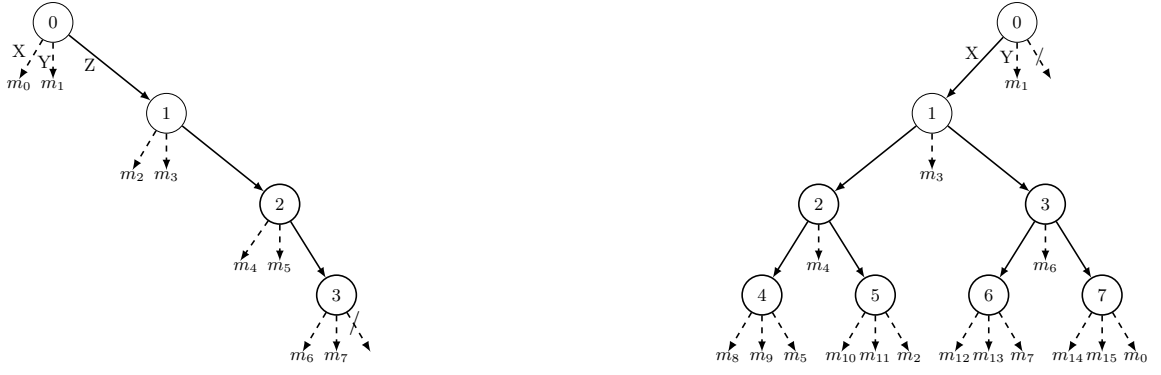

\subsection{Jordan--Wigner and Bravyi--Kitaev mappings}

The Jordan--Wigner (JW) transformation associates the occupation number of the $i$-th fermionic mode with the computational basis state of the $i$-th qubit, corresponding to the identity mapping $J$. The Majorana operators are represented as
\begin{equation}
\begin{aligned}
    f_{\text{JW}}(m_{2i}) &= X_i \prod_{j=0}^{i-1} Z_j, \\
    f_{\text{JW}}(m_{2i+1}) &= Y_i \prod_{j=0}^{i-1} Z_j.
\end{aligned}
\end{equation}
In this mapping, the parity of fermionic mode $i$ is encoded in the $Z$ operators acting on the preceding qubits, resulting in a Pauli weight that scales linearly with the system size, $\mathcal{O}(n)$. The corresponding tree structure is shown in Fig. \ref{fig:TT_examples}.

The Bravyi--Kitaev (BK) transformation organizes fermionic occupation and parity information using a binary tree structure \cite{Bravyi,Seeley_2012}. This structure can be rearranged into a ternary tree representation, as illustrated in Fig. \ref{fig:TT_examples}. By storing partial sums of occupation numbers across different qubits, the BK mapping achieves a Pauli weight scaling of $\mathcal{O}(\log_2 n)$.

\subsection{Bonsai algorithm }
\label{Sec:Bonsai}
A fermion-to-qubit mapping tailored to specific hardware connectivity graphs can be constructed using the Bonsai algorithm introduced in \cite{Miller}. Given a connectivity graph $\mathcal{G}$, the algorithm builds a ternary tree $\mathcal{T}=(V_{\mathcal{T}},L_{\mathcal{T}})$ according to the following procedure:
\begin{enumerate}
    \item Initialize $V_{\mathcal{T}}=\emptyset$ and $L_{\mathcal{T}}=\emptyset$.
    \item Select a root node $r$ such that
    \[
        r=\arg\min_u \max_v d(u,v),
    \]
    where $d(u,v)$ denotes the topological distance between nodes $u$ and $v$ in $\mathcal{G}$.
    \item Define the first level $\mathcal{L}_0=\{r\}$ and construct $\mathcal{L}_1$ by attaching up to three neighboring nodes of $r$.
    \item For each subsequent level $\mathcal{L}_k$, attach nodes not yet included in $V_{\mathcal{T}}$ to the nodes in $\mathcal{L}_k$, forming $\mathcal{L}_{k+1}$.
    \item Repeat the procedure until the entire graph $\mathcal{G}$ is spanned.
\end{enumerate}

Once the tree is generated, legs are added to nodes with fewer than three downward links to complete the ternary structure. The resulting tree is labeled using the \emph{homogeneous localization} scheme described in \cite{Miller}, in which the Pauli operators $X$, $Y$, and $Z$ are assigned to downward links in a fixed priority order, with remaining labels assigned to legs. An alternative \emph{heterogeneous localization} strategy, also proposed in \cite{Miller}, localizes certain fermionic modes onto fewer qubits at the expense of delocalizing others, potentially leading to deeper circuits. Since no specific assumptions can be made about the structure of the physical system or the hardware, homogeneous localization is adopted throughout this work.

\begin{figure*}[t]
    \centering

    \begin{minipage}[t]{0.45\textwidth}    
        \vspace{0pt}
        \centering
        \scalebox{1.5}{
\begin{tikzpicture}[
    scale=1, 
    every node/.style={font=\tiny}, 
    dot/.style={circle, draw=black, fill=white, minimum size=4mm, inner sep=0pt}, 
    dotOrange/.style={dot, fill=orange!30},
    dotBlue/.style={dot, fill=cyan!30},
    conn/.style={gray!80, line width=0.8pt}
]

\coordinate (V0) at (90:1.2);   
\coordinate (V1) at (30:1.2);   
\coordinate (V2) at (-30:1.2);  
\coordinate (V3) at (-90:1.2);  
\coordinate (V4) at (210:1.2);  
\coordinate (V5) at (150:1.2);  

\coordinate (E0) at (60:1.04);  
\coordinate (E1) at (0:1.04);   
\coordinate (E2) at (-60:1.04); 
\coordinate (E3) at (-120:1.04);
\coordinate (E4) at (180:1.04); 
\coordinate (E5) at (120:1.04); 

\coordinate (A0) at (90:1.7);   
\coordinate (A1) at (30:1.7);   
\coordinate (A2) at (-30:1.7);  
\coordinate (A3) at (-90:1.7);  
\coordinate (A4) at (210:1.7);  
\coordinate (A5) at (150:1.7);  

\draw[conn] (V0)--(E0)--(V1)--(E1)--(V2)--(E2)--(V3)--(E3)--(V4)--(E4)--(V5)--(E5)--(V0);
\foreach \i in {0,...,5} { \draw[conn] (V\i) -- (A\i); }

\node[dotBlue] at (V0) {14};
\node[dotOrange] at (E0) {15};
\node[dotOrange] at (V1) {13};
\node[dotOrange] at (E1) {9};
\node[dotOrange] at (V2) {7};
\node[dotOrange] at (E2) {3};
\node[dotOrange] at (V3) {1};
\node[dot] at (E3) {0};
\node[dotBlue] at (V4) {2};
\node[dotBlue] at (E4) {6};
\node[dotBlue] at (V5) {8};
\node[dotBlue] at (E5) {12};

\node[dotBlue] at (A0) {17};
\node[dotOrange] at (A1) {16};
\node[dotOrange] at (A2) {10};
\node[dotOrange] at (A3) {4};
\node[dotBlue] at (A4) {5};
\node[dotBlue] at (A5) {11};

\end{tikzpicture}}
\end{minipage}
\hfill
    \begin{minipage}[t]{0.45\textwidth}    
        \vspace{0pt}
        \centering
        \scalebox{1}{
     \begin{tikzpicture}[
            level distance=0.8cm,
            sibling distance=1cm,
            every node/.style={draw,circle,minimum size=4mm,inner sep=0pt,font=\tiny},
            edge from parent/.style={draw,-latex,thick},
            leg/.style={draw,-latex,dashed,thick},
            dot/.style={circle, draw=black, fill=white, minimum size=4mm, inner sep=0pt}, 
            dotOrange/.style={dot, fill=orange!30},
            dotBlue/.style={dot, fill=cyan!30},
        ]
\node (r) {0}
    child [sibling distance=2cm] {
        node [dotOrange](n1) {1}    
        child [sibling distance=1cm] {
            node [dotOrange] (n3) {3}     
            child [sibling distance=0.5cm] {
                node [dotOrange](n7) {7}
                child {
                    node [dotOrange](n9) {9}
                    child [sibling distance=1cm] {
                        node [dotOrange](n13) {13}
                        child { node [dotOrange](n15) {15} }
                        child { node[dotOrange] (n16) {16} }
                    }
                }
                child [sibling distance=1.8cm]{ node [dotOrange](n10) {10} }
            }
            child[missing]
        }
        child [sibling distance=1cm] {
            node [dotOrange] (n4) {4}
        }
    }
    child [sibling distance=2cm] {
        node [dotBlue] (n2) {2} 
        child [sibling distance=1cm] {
            node [dotBlue](n5) {5} 
        }
        child [sibling distance=1cm] {
            node [dotBlue] (n6) {6} 
            child[missing]
            child [sibling distance=0.5cm] {
                node [dotBlue] (n8) {8}
                child [sibling distance=2.5cm] { node [dotBlue] (n11) {11} }
                child {
                    node [dotBlue](n12) {12}
                    child [sibling distance=1cm] {
                        node[dotBlue] (n14) {14}
                        child { node [dotBlue] (n17) {17} }
                    }
                }
            }
        }
    };


\draw[leg] (r) -- ++(0.0,-0.6);
\draw[leg] (n1) -- ++(1.0,-0.6);
\draw[leg] (n2) -- ++(0.0,-0.6);
\draw[leg] (n7) -- ++(1.4,-0.6);
\draw[leg] (n13) -- ++(1.4,-0.6);

\draw[leg] (n3) -- ++(0.15,-0.6);   
\draw[leg] (n3) -- ++(0.4,-0.6);   

\draw[leg] (n6) -- ++(-0.4,-0.6);  
\draw[leg] (n6) -- ++(-0.15,-0.6);   

\draw[leg] (n8) -- ++(-0.4,-0.6);   

\draw[leg] (n9) -- ++(+0.3,-0.6);   
\draw[leg] (n9) -- ++(+0.6,-0.6);  

\draw[leg] (n12) -- ++(-0.6,-0.6); 
\draw[leg] (n12) -- ++(-0.3,-0.6);  

\draw[leg] (n14) -- ++(-0.6,-0.6);  
\draw[leg] (n14) -- ++(-0.3,-0.6);  

\foreach \i in {n4, n5, n10, n11, n15, n16, n17} {
    \draw[leg] (\i) -- ++(-0.4,-0.6);
    \draw[leg] (\i) -- ++(0.0,-0.6);
    \draw[leg] (\i) -- ++(0.4,-0.6);
}
           
        \node[font=\small, draw=none] at ($(r)!0.33!(a)+(-0.2,0.3)$) {X};
        \node[font=\small, draw=none] at ($(r)+(-0.18,-0.6)$) {Y};

        \end{tikzpicture}}
    \end{minipage}
    \caption{The Bonsai mapping for the 18 qubit chip. First, we span the entire heavy-hexagon connectivity graph to build the tree structure as illustrated in Subsec. \ref{Sec:Bonsai}. Then, we use homogeneous localization to associate the $X$,$Y$ and $Z$ Pauli gates to the legs and edges properly. Note that in the first branch the priority order of labeling is $X,Y,Z$, while in the second branch $Z,X,Y$ in order to maximize the length of the path with only $Z$s, as described in \cite{Miller}.}
    \label{fig:bonsai}
\end{figure*}
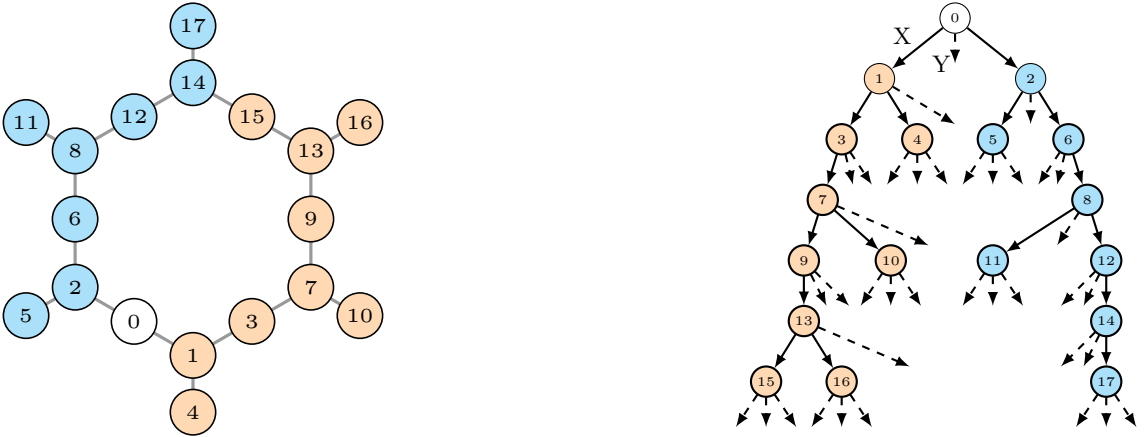

As an example, let us consider the building block of the Heron r3 IBM heavy-hexagon quantum processor, which is here reduced to 18 qubit for simplicity.
The final mapping obtained through this procedure is illustrated in Fig.~\ref{fig:bonsai}.
The Pauli weight resulting from the Majorana strings of this greedy algorithm is $\mathcal{O}(\sqrt{n})$, because the strings are proportional to a radius $R$ of the area of the chip and number of qubits $n$ at a given topological distance  smaller than $R$ scales as $R^2$ \cite{Miller}.

\subsection{Time evolution of many-fermion systems}
\label{Sec: decomposition}

For simulating the time evolution of a many particle system we must exponentiate the associated Hamiltonian and split, via Trotter-like decompositions, the unitary into a product of implementable quantum gates \cite{Campbell_2019,suzuki,suzuki2,Berry_2006}. These are typically a subset of single- and two-qubit gates. For the heavy-hexagon architecture these are 
$Sx,X,R_x,R_z,Rzz,\mathrm{CZ}$. 
Each Trotter step belongs to three possible cases:
\begin{itemize}
    \item \textit{Strings with only one Pauli operator:}
if only one Pauli operator is exponentiated, the decomposition is trivial, as it is sufficient to implement an $R_x$ or $R_z$ rotation for the $X$ and $Z$ gates respectively. In the case of the $Y$ gate, we need to perform a change of basis, as $R_y(\theta)=R_z(\frac{\pi}{2})\cdot R_x(\theta)\cdot R_z(-\frac{\pi}{2}).$
    \item \textit{Strings with two Pauli operators:}
if the transpiler encounters an exponential of a string with two Paulis, it simply implements the the native $Rzz$ gates as it collects less noise that two $\mathrm{CZ}$ gates. If the string is $Z \otimes Z$, we can directly use $Rzz$, instead, if we have the $X$ and $Y$ operators, a change of basis is required, as previously, $ X=H\cdot Z\cdot H$, $ Y=R_x(-\frac{\pi}{2})\cdot Z\cdot R_x(\frac{\pi}{2}), $ with $H=R_z(\frac{\pi}{2})\cdot R_x(\frac{\pi}{2})\cdot R_z(\frac{\pi}{2})$.
    \item \textit{Strings with more than two Pauli operators: }
let us consider a string composed only of $Z$ operators, as otherwise it would require just a change of basis, as before. The decomposition is applied in such a manner that the parity of the qubits is stored in the last one through $\mathrm{CX}$ (\text{\footnotesize CNOT}) gates \cite{Seeley_2012}:
\begin{align}
      \exp(i \theta Z_0 \dots Z_n)&= \mathrm{CX}_{0,1}\cdot \mathrm{CX}_{1,2} \dots \mathrm{CX}_{n-2,n-1} \label{decomp}
                               \\ & \cdot Rzz_{n-1,n}(2\theta)\cdot \mathrm{CX}_{n-2,n-1} \dots \mathrm{CX}_{0,1}, \nonumber
\end{align}
with $\mathrm{CX}_{i,j}$=$H_j \cdot \mathrm{CZ}_{i,j} \cdot H_j$.
\end{itemize}

As we notice, the operations involved are local, so if we consider the two qubit gates, they operate on two qubits that are near to each other. 
Now, if the Pauli string acts on a set of non-contiguous qubits on the physical hardware, it is necessary to get them closer through additional \text{\footnotesize SWAP} gates. Naturally, as \text{\footnotesize SWAP} operations consists of three $\mathrm{CZ}$ gates, they are generally expensive in terms of elementary gates, and as a consequence, the scalability of a given transformation could dramatically increase, making that mapping very inefficient on a real quantum processor. This feature will be further explained with the numerical results found in the real scalability for heavy-hexagon quantum processors in the following Section \ref{section:classical_results}.

\section{Classical benchmarks for fermion-to-qubit transformations}
\label{section:classical_results}

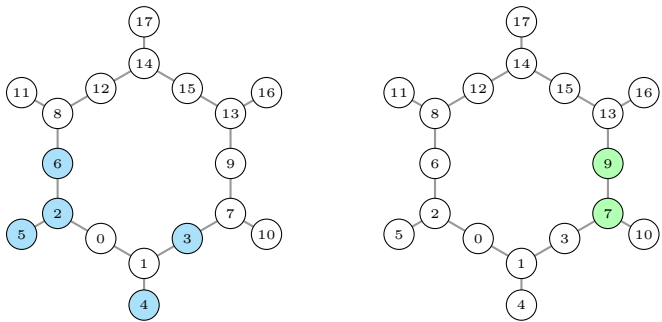
\begin{figure}
   \centering
\begin{tikzpicture}[
    scale=1.1, 
    every node/.style={font=\tiny}, 
    dot/.style={circle, draw=black, fill=white, minimum size=4mm, inner sep=0pt}, 
    dotBlue/.style={dot, fill=cyan!30},
    conn/.style={gray!80, line width=0.8pt}
]

\coordinate (V0) at (90:1.2);   
\coordinate (V1) at (30:1.2);   
\coordinate (V2) at (-30:1.2);  
\coordinate (V3) at (-90:1.2);  
\coordinate (V4) at (210:1.2);  
\coordinate (V5) at (150:1.2);  

\coordinate (E0) at (60:1.04);  
\coordinate (E1) at (0:1.04);   
\coordinate (E2) at (-60:1.04); 
\coordinate (E3) at (-120:1.04);
\coordinate (E4) at (180:1.04); 
\coordinate (E5) at (120:1.04); 

\coordinate (A0) at (90:1.7);   
\coordinate (A1) at (30:1.7);   
\coordinate (A2) at (-30:1.7);  
\coordinate (A3) at (-90:1.7);  
\coordinate (A4) at (210:1.7);  
\coordinate (A5) at (150:1.7);  

\draw[conn] (V0)--(E0)--(V1)--(E1)--(V2)--(E2)--(V3)--(E3)--(V4)--(E4)--(V5)--(E5)--(V0);
\foreach \i in {0,...,5} { \draw[conn] (V\i) -- (A\i); }

\node[dot] at (V0) {14};
\node[dot] at (E0) {15};
\node[dot] at (V1) {13};
\node[dot] at (E1) {9};
\node[dot] at (V2) {7};
\node[dotBlue] at (E2) {3};
\node[dot] at (V3) {1};
\node[dot] at (E3) {0};
\node[dotBlue] at (V4) {2};
\node[dotBlue] at (E4) {6};
\node[dot] at (V5) {8};
\node[dot] at (E5) {12};

\node[dot] at (A0) {17};
\node[dot] at (A1) {16};
\node[dot] at (A2) {10};
\node[dotBlue] at (A3) {4};
\node[dotBlue] at (A4) {5};
\node[dot] at (A5) {11};

\end{tikzpicture}
\hfill 
\begin{tikzpicture}[
    scale=1.1, 
    every node/.style={font=\tiny}, 
    dot/.style={circle, draw=black, fill=white, minimum size=4mm, inner sep=0pt}, 
    dotGreen/.style={dot, fill=green!30},
    conn/.style={gray!80, line width=0.8pt}
]

\coordinate (V0) at (90:1.2);   
\coordinate (V1) at (30:1.2);   
\coordinate (V2) at (-30:1.2);  
\coordinate (V3) at (-90:1.2);  
\coordinate (V4) at (210:1.2);  
\coordinate (V5) at (150:1.2);  

\coordinate (E0) at (60:1.04);  
\coordinate (E1) at (0:1.04);   
\coordinate (E2) at (-60:1.04); 
\coordinate (E3) at (-120:1.04);
\coordinate (E4) at (180:1.04); 
\coordinate (E5) at (120:1.04); 

\coordinate (A0) at (90:1.7);   
\coordinate (A1) at (30:1.7);   
\coordinate (A2) at (-30:1.7);  
\coordinate (A3) at (-90:1.7);  
\coordinate (A4) at (210:1.7);  
\coordinate (A5) at (150:1.7);  

\draw[conn] (V0)--(E0)--(V1)--(E1)--(V2)--(E2)--(V3)--(E3)--(V4)--(E4)--(V5)--(E5)--(V0);
\foreach \i in {0,...,5} { \draw[conn] (V\i) -- (A\i); }

\node[dot] at (V0) {14};
\node[dot] at (E0) {15};
\node[dot] at (V1) {13};
\node[dotGreen] at (E1) {9};
\node[dotGreen] at (V2) {7};
\node[dot] at (E2) {3};
\node[dot] at (V3) {1};
\node[dot] at (E3) {0};
\node[dot] at (V4) {2};
\node[dot] at (E4) {6};
\node[dot] at (V5) {8};
\node[dot] at (E5) {12};

\node[dot] at (A0) {17};
\node[dot] at (A1) {16};
\node[dot] at (A2) {10};
\node[dot] at (A3) {4};
\node[dot] at (A4) {5};
\node[dot] at (A5) {11};
\end{tikzpicture}

\caption{An illustrative comparison between JW and Bonsai mapping. Both the transformations implement the fermionic single excitation $m_4m_{12}$. The coloured nodes are the actual qubits on which the single excitation operates. Due to Pauli cancellations, the qubits on which the JW mapping acts are disconnected, and a \text{\footnotesize SWAP} operation on qubit $0$ and $1$ is necessary to reconnect all the nodes. By contrast, the Bonsai mapping does not require extra operations. Note that for a chip of this size in principle we could label the nodes for the JW mapping in order to reduce the overhead operations, but in practice, for a larger graph, the \text{\footnotesize SWAP} is determined by the topology, independently from the labeling chosen for the qubits.}
    \label{fig:swap}
\end{figure}
We consider three benchmark fermion-to-qubit mappings, Jordan-Wigner (JW), Bravyi-Kitaev (BK), and the Bonsai transformation,  applied to IBM heavy-hexagon quantum processors, which provide a paradigmatic example of constrained superconducting-qubit connectivity. 
A representative example illustrating the impact of hardware connectivity constraints on JW and Bonsai mappings is shown in Fig.~\ref{fig:swap}. 
The benchmark focuses on single and double Majorana excitations, corresponding to the terms appearing in the Hubbard Hamiltonian.

Importantly, we notice that the enumeration of the qubits on hardware graphs with more than one or two hexagonal cells does not matter, as the efficiency of the mapping is mostly determined by the topology of the underlying graph. For this reason, the qubit labels adopted throughout this section follow the ordering prescribed by the Bonsai procedure.
We importantly remark that, even if for the first and the second chip considered in our work ($18$ and $27$ qubits) a strategic choice for the labeling of the qubits may reduce the overall routing for $JW$ and $BK$ mappings, these yield only a marginal advantage over our configuration, and in any case does not affect the results registered for larger chips and the analysis done in this work.

In the Jordan--Wigner transformation, fermionic operators are mapped to Pauli strings that are generically nonlocal and typically span the entire hardware graph. 
As a consequence, the mapping overhead should also scale linearly with system size. 
The Bravyi--Kitaev transformation reduces the Pauli support by encoding parity information more efficiently; however, when implemented on hardware with restricted connectivity, the resulting Pauli strings are often supported on disconnected subsets of qubits. 
Restoring locality in this case requires additional \text{\footnotesize SWAP} operations, causing the ideal logarithmic scaling to be lost once routing is explicitly taken into account~\cite{Miller}. 
Finally, the Bonsai transformation is specifically designed to match the underlying hardware topology. 
As a result, no \text{\footnotesize SWAP} operations are required for single excitations, while only a limited routing overhead is expected for double excitations.

In this work, we perform a numerical analysis of the effective scalability of these three mappings across heavy-hexagon processors of increasing size, explicitly accounting for the routing overhead induced by hardware constraints. 
We compute the average Pauli weight for single and double excitations and study their asymptotic behaviour. 
A further complication arises in estimating the number of required \text{\footnotesize SWAP} operations. 
Given a Pauli string supported on a disconnected subset of qubits, determining the minimal routing cost reduces to a Steiner tree problem on the hardware graph, i.e., finding the shortest subgraph connecting all involved qubits. 
Since this problem is NP-hard \cite{Steger2002Steiner}, an exact solution rapidly becomes impractical even for systems comprising a few tens of qubits.

To address this issue, we implement a heuristic approach inspired by the Takahashi--Matsuyama (TM) algorithm \cite{Takahashi1980Steiner}.
More sophisticated Steiner-tree heuristics exist that provide improved accuracy and runtime for general graphs~\cite{shrub}. 
Nevertheless, the TM-inspired approach is particularly well suited to the heavy-hexagon topology, where shortest paths are strongly constrained and the heuristic solution often coincides with the exact one. We briefly discuss the functioning of this heuristic.
First, contiguous subsets of qubits are grouped into blocks, within which no \text{\footnotesize SWAP} operations are required. 
These blocks are then contracted into representative nodes, yielding a reduced graph that preserves the original connectivity outside the blocks. 
On this reduced graph, a minimum spanning tree is constructed by iteratively connecting the pair of terminals with minimal topological distance, producing an approximate Steiner tree.

The difference $\epsilon$ between $L_{\rm TM}$ and a fixed $L_{\rm opt}$, i.e., the number of nodes linking together the terminals found by the approximate and the exact solution to the Steiner tree problem, cannot be more than
\begin{equation}
  \epsilon \leq L_{\rm opt}\left(1-\frac{2}{T}\right),\quad T\ge 2,
    \label{bound}
\end{equation}
where $T$ is the number of terminals to be connected \cite{Takahashi1980Steiner}.
The procedure is presented in more detail in Algorithm~\ref{alg:swap_count} in Appendix \ref{app:TM-alg}.
However, Eq.~(\ref{bound}) considers only the theoretical error bound associated to $L_{\rm opt}$, that is, given $L_{\rm opt}$, the $TM$ solution must sit in a neighborhood $\epsilon$ bounded by Eq.~(\ref{bound}). 
In our setting, the bound must be expressed in terms of the approximate value $L_{\rm TM}$ rather than $L_{\rm opt}$, and, consequently, an appropriate error must be derived. Starting from Eq.~(\ref{bound}) and using $\epsilon = L_{\rm TM}-L_{\rm opt}$, 
the resulting bound on the approximation error for $T> 2$ reads
\begin{equation}
  \epsilon \leq \frac{L_{\rm TM}}{2} \frac{T-2}{T-1} \leq \frac{L_{\rm TM}}2.
  \label{eq:eps upper}
\end{equation}

\begin{table}[t]
    \centering
    \begin{tabular}{l l l l l l}
        \toprule
        Mapping & $N_{q}$ & $\omega(S_{s})$ & $\omega(S_{d})$ & \text{\footnotesize SWAP}($S_{s}$) & \text{\footnotesize SWAP}($S_{d}$) \\
        \midrule
        Jordan-Wigner & 18 & 14.30 & 18.52 & 5.9 $\pm^{0}_{1.1}$ & 8.028 $\pm^{0}_{2.4}$ \\
        ~ & 27 & 20.31 & 25.81 & 15.5 $\pm^{0}_{5.1}$ & 16.5 $\pm^{0}_{6.2}$ \\
        ~ & 37 & 26.99 & 33.86 & 20.15 $\pm^{0}_{7.4}$ & 21.2 $\pm^{0}_{8.6}$ \\
        ~ & 53 & 37.66 & 46.71 & 23.1 $\pm^{0}_{8.8}$ & 28 $\pm^{0}_{12}$ \\
        ~ & 65 & 45.66 & 56.32 & 27 $\pm^{0}_{11}$ & 34 $\pm^{0}_{15}$ \\
        ~ & 99 & 68.33 & 83.48 & 34 $\pm^{0}_{14}$ & 45 $\pm^{0}_{20}$ \\
        ~ & 133 & 91.00 & 110.8 & 42 $\pm^{0}_{18}$ & 58 $\pm^{0}_{27}$ \\
        \addlinespace

        Bravyi-Kitaev & 18 & 10.30 & 15.77 & 9.8 $\pm^{0}_{2.7}$ & 11.1 $\pm^{0}_{3.7}$ \\
        ~ & 27 & 12.52 & 19.17 & 17.5 $\pm^{0}_{5.5}$ & 19.2 $\pm^{0}_{7.2}$ \\
        ~ & 37 & 13.65 & 21.78 & 28 $\pm^{0}_{11}$ & 31 $\pm^{0}_{13}$ \\
        ~ & 53 & 15.93 & 25.36 & 37 $\pm^{0}_{15}$ & 42 $\pm^{0}_{18}$ \\
        ~ & 65 & 16.58 & 26.84 & 46 $\pm^{0}_{18}$ & 52 $\pm^{0}_{23}$ \\
        ~ & 99 & 19.41 & 31.70 & 64 $\pm^{0}_{27}$ & 76 $\pm^{0}_{34}$ \\
        ~ & 133 & 20.32 & 34.01 & 88 $\pm^{0}_{38}$ & 107 $\pm^{0}_{49}$ \\
        \addlinespace

        Bonsai & 18 & 12.23 & 16.88 & - & 3.431 $\pm^{0}_{0.034}$ \\
        ~ & 27 & 14.30 & 21.31 & - & 4.106 $\pm^{0}_{0.028}$ \\
        ~ & 37 & 16.41 & 24.95 & - & 4.604 $\pm^{0}_{0.014}$ \\
        ~ & 53 & 19.67 & 30.41 & - & 5.606 $\pm^{0}_{0.011}$ \\
        ~ & 65 & 21.97 & 34.21 & - & 6.229 $\pm^{0}_{0.015}$ \\
        ~ & 99 & 26.51 & 42.24 & - & 7.255 $\pm^{0}_{0.011}$ \\
        ~ & 133 & 31.39 & 50.38 & - & 8.738 $\pm^{0.009}_{0.012}$ \\
        \bottomrule
    \end{tabular}
    \caption{Values of the Pauli weights $\omega$ and the \text{\footnotesize SWAP} operations of single ($S_s$) and double ($S_d$) excitations strings,  introduced in Eq.~(\ref{eq:pauli_weight}), reported in terms of the number of equivalent $\mathrm{CZ}$ gates applied for a heavy-hexagon topology.
    The Bonsai mapping registered the lower number of $\mathrm{CZ}$s above all the transformations.}
    \label{tab:mappature}
\end{table}

\begin{table}[t]
    \centering
    \begin{tabular}{llll}
    \toprule
     & c & p & \text{\footnotesize RMS}$_{rel}$ \\ 
    \midrule
    $\boldsymbol{\omega(S_s)}$ &  &  & \\ 
    Jordan-Wigner & 1.09 ± 0.06 & 0.907 ± 0.012 & 1.44\% \\ 
    Bravyi-Kitaev (log)  & 1.30 ± 0.12 & 1.39 ± 0.05 & 1.65\% \\ 
    Bonsai & 3.51 ± 0.17 & 0.454 ± 0.011 & 1.67\% \\ 
    \addlinespace
    
    $\boldsymbol{\omega(S_d)}$ & ~ & ~ & \\ 
    Jordan-Wigner & 1.44 ± 0.08 & 0.890 ± 0.013 & 1.60\% \\ 
    Bravyi-Kitaev (log) & 1.89 ± 0.13 & 1.48 ± 0.03 & 0.95\% \\ 
    Bonsai & 4.08 ± 0.10 & 0.518 ± 0.006 & 0.81\% \\ 
    \addlinespace
    
    \textbf{\footnotesize SWAP}$\boldsymbol{(S_s)}$ &  &  & \\
    Jordan-Wigner & 0.631 ± 0.006 & 0.8671 ± 0.0019 & 19.69\%\\
    Bravyi-Kitaev & 0.650 ± 0.004 & 1.0046 ± 0.0015 & 7.91\% \\ 
    \addlinespace
    
   \textbf{\footnotesize SWAP}$\boldsymbol{(S_d)}$ & ~ & ~ & \\ 
    Jordan-Wigner & 0.736 ± 0.005 & 0.8995 ± 0.0015 & 10.34\% \\ 
    Bravyi-Kitaev & 0.646 ± 0.004 & 1.0424 ± 0.0014 & 6.28\% \\ 
    Bonsai & 1.51 ± 0.04 & 0.377 ± 0.006 & 4.07\% \\ 
    \bottomrule
    \end{tabular}
    \caption{Slope and exponent obtained from the polynomial and log-polynomial laws. Consistently with the scaling of the string directly generated from the mappings, JW single and double excitations are sub-linear, BK logarithmic and Bonsai subquadratic. }
    \label{tab:fit_model}
\end{table}

Based on these considerations, in Tab.~\ref{tab:mappature} we report the average Pauli weight and routing overhead for single and double excitations, already expressed in units of $\mathrm{CZ}$-gate cost.
For the largest processors (99 and 133 qubits), the reported values are obtained from a representative sampling of excitations, as an exhaustive enumeration becomes computationally prohibitive.

The results show that the Pauli weights of single and double excitations follow the asymptotic scaling of the underlying Majorana strings. 
When routing costs are included, the JW mapping exhibits a substantial overhead originating from the nonlocal structure of its Pauli strings, which fragment over multiple disconnected regions of the hardware. 
The BK transformation achieves the smallest Pauli weights among the three mappings; however, this advantage is completely offset by the large number of \text{\footnotesize SWAP} operations required to restore locality, in agreement with previous observations~\cite{Miller}. 
By contrast, the Bonsai mapping incurs no routing overhead for single excitations and only a modest cost for double excitations, resulting in a significantly reduced overall circuit depth.

We now observe from the numerical data in Tab.~\ref{tab:mappature} that the relative accuracy of the TM approximation deteriorates as the number of nodes in the hardware graph increases, approaching the asymptotic $50\%$ limit imposed by the theoretical bound Eq.~\eqref{eq:eps upper}.
Since the relative error depends solely on the number of terminals $T$, this behavior reflects the structural properties of the mapped operators on the specific graph geometry.

For the JW and BK mappings, the data indicate that the large routing overhead is primarily associated with a high number of terminals distributed across the graph. 
In this regime, the heuristic is required to connect many terminals that are mutually close in graph distance but belong to distinct disconnected blocks, resulting in numerous local Steiner connections and, consequently, a large number of \text{\footnotesize SWAP} operations. 
The degradation of accuracy toward the $50\%$ limit is therefore consistent with a fragmentation pattern characterized by many nearby terminals rather than by a few widely separated ones.

By contrast, the Bonsai mapping exhibits a markedly different behavior: the corresponding Steiner trees involve only a small number of terminals, leading both to a tighter relative accuracy and to a significantly reduced routing overhead.

\begin{figure*}[t]
    \centering
\includegraphics[width=2.0\columnwidth]{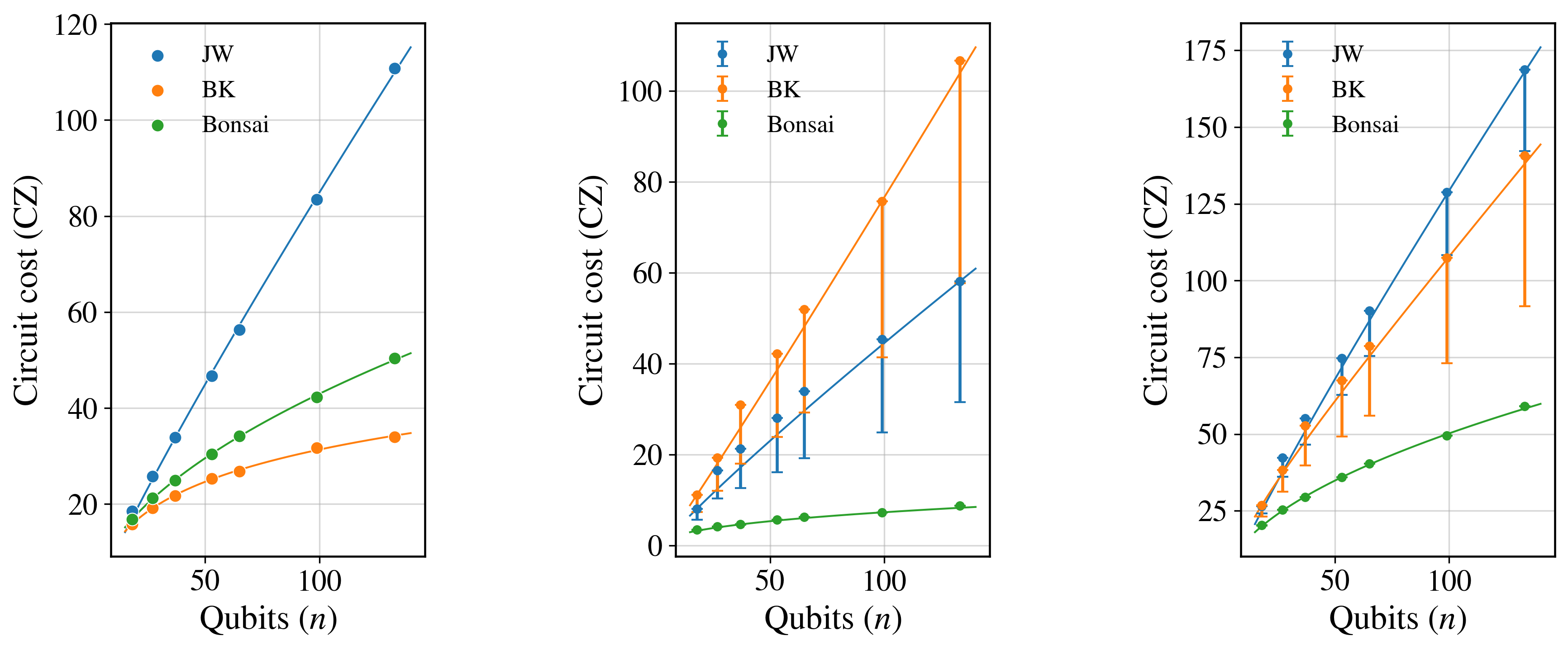}
          
\caption{On the left, the ideal Pauli weight scaling of the $m_{i}m_{j}m_{k}m_{l}$ interactions, expressed in terms of the number of $\mathrm{CZ}$ gates ($y$-axis) to be executed for JW, BK and Bonsai mappings, as a function of the number of qubits ($x$-axis). At the center, the additional \text{\footnotesize SWAP} operations (isolated from ideal Pauli weight of the left picture) required to implement these interactions  on chips with heavy-hexagon connectivity without violating the locality principle, quantified by the number of $\mathrm{CZ}$ to be perform, in order to make the comparison between the two figures more readable. On the right, the composition of the first two graph, in order to show the effective circuit cost of the three maps on a heavy-hexagon graph.
The curves depicted in the figure are obtained by fitting  the mean values on the all possible $m_{i}m_{j}m_{k}m_{l}$ combinations for seven different processors (except for the chips with $99$ and $133$ qubits, where we take a large representative sampling). The data computed are represented by dots.
Finally, the error bound associated to the routing operations takes into account that all the combinations computed deviate from the optimal Steiner Tree by the maximum possible quantity, as described in Eq. \ref{eq:eps upper}. Consequently, the actual error bound is therefore substantially tighter than the one shown.}
    \label{fig:fit_graphs}
\end{figure*}

We fit the average Pauli weight for single and double excitations using the functional form $c\,n^{p}+d$ for the JW and Bonsai mappings, and $c\,\log_2^{p}(n)+d$ for the BK transformation. 
This choice is motivated by the fact that single and double excitations correspond to products of two or four Majorana strings, whose asymptotic scaling is analytically known \cite{JW,Bravyi,Miller}. 
Pauli cancelation effects may therefore introduce only subleading deviations from the expected scaling behavior.
Moreover, based on the exact results for $n=2$ or $n=1$ qubits, it is possible to fix the $d$ coefficient in order to reduce the d.o.f.~of the total fit.

The fitted exponents are recorded in Tab.~\ref{tab:fit_model}. Through the numerical data and the fit registered, we plot the mapping scalability in Fig.~\ref{fig:fit_graphs}. 
The extracted scalings confirm that JW and Bonsai single and double excitations exhibit sublinear and subquadratic behavior, respectively, due to Pauli cancelation effects. 
In the BK case, the effective Pauli weight increases relative to the ideal mapping as a consequence of the spatial dispersion of BK qubits across the hardware graph.

Finally, the larger root-mean-square deviations (\text{\footnotesize RMS}$_{rel}$) observed for the routing fits---particularly for JW single excitations---can be attributed to variations in the detailed structure of different heavy-hexagon processors. 
Differences in the number of hexagons and symmetry axes affect the spanning-tree construction and, consequently, the routing cost. 
This sensitivity is especially pronounced for mappings that generate highly nonlocal Pauli strings and suggests an interesting direction for further quantitative investigation.

\section{Simulating the 2D Hubbard model}
\label{Sec: simulation}

Having established numerically that the Bonsai transformation provides superior performance with respect to Jordan--Wigner and Bravyi--Kitaev mappings on heavy-hexagon hardware, we now investigate how additional knowledge of the physical system can be exploited to further optimize the fermion-to-qubit encoding. In particular, we focus on the simulation of the two-dimensional Hubbard model and on the impact of mapping-dependent circuit costs on near-term quantum devices.

\subsection{Time evolution}
\label{sec:hamiltonian}

The physical system under consideration is a two-dimensional $L\times L$ spinful Fermi-Hubbard model on a square lattice.
The corresponding Hamiltonian can be written in terms of Majorana operators as
\begin{equation}
\begin{aligned}
\hat{H} = &  -\frac{J}{2} \sum_{\langle i,j\rangle,\sigma} i
\left(
m_{2i,\sigma} m_{2j+1,\sigma}
- m_{2i+1,\sigma} m_{2j,\sigma}
\right) \\ &
- \frac{U}{4}  \sum_i 
\Big[ i \left(
m_{2i+1,\uparrow} m_{2i,\uparrow}
+ m_{2i+1,\downarrow} m_{2i,\downarrow} \right)
\\ & + m_{2i+1,\uparrow} m_{2i,\uparrow}
  m_{2i+1,\downarrow} m_{2i,\downarrow}
\Big],
\end{aligned}
\label{eq:hamiltonian}
\end{equation}
where $i$ and $j$ are the label lattice sites, from 1 to $L$, and $\sigma\in\{\uparrow,\downarrow\}$ denotes the spin degree of freedom. In this work we mostly use $L=4$ or $L=6$. 

The key ingredient for quantum simulation is the implementation of the unitary time evolution. As is well known, quantum states evolve according to the Schr\"odinger equation, $\ket{\psi(t)} = e^{-i \hat{H} t}\ket{\psi(0)}.$ From a quantum-computing perspective, the operator $e^{-i \hat{H} t}$ must be decomposed into a sequence of elementary gates compatible with the target hardware.

In this work, we employ the qDRIFT protocol \cite{Campbell_2019,david2025tightererrorboundsqdrift} for two main reasons.
First, the simulated evolution times are relatively short, a regime in which the error bounds of qDRIFT are expected to be tighter than those of Trotter--Suzuki decompositions \cite{Campbell_2019,suzuki,suzuki2,Berry_2006}. Second, the performance of qDRIFT improves in the presence of long-range interaction terms, which play a crucial role in determining the physical properties of fermionic lattice models \cite{Campbell_2019}. This feature makes the approach naturally extensible to more complex Hamiltonians.
While better asymptotic precision could be obtained via Quantum Signal Processing techniques \cite{low2017optimal,motlagh2024generalized}, these would require many ancillary qubits, making the hardware simulation on current hardware much more complicated. 
The qDRIFT protocol, together with the improved error bound of \cite{david2025tightererrorboundsqdrift} and the details about the critical parameters in our simulation, are studied in depth in Appendix~\ref{app:qDRIFT protocol}.

\begin{figure}[t]
  \centering
  \includegraphics[width=0.5\textwidth]{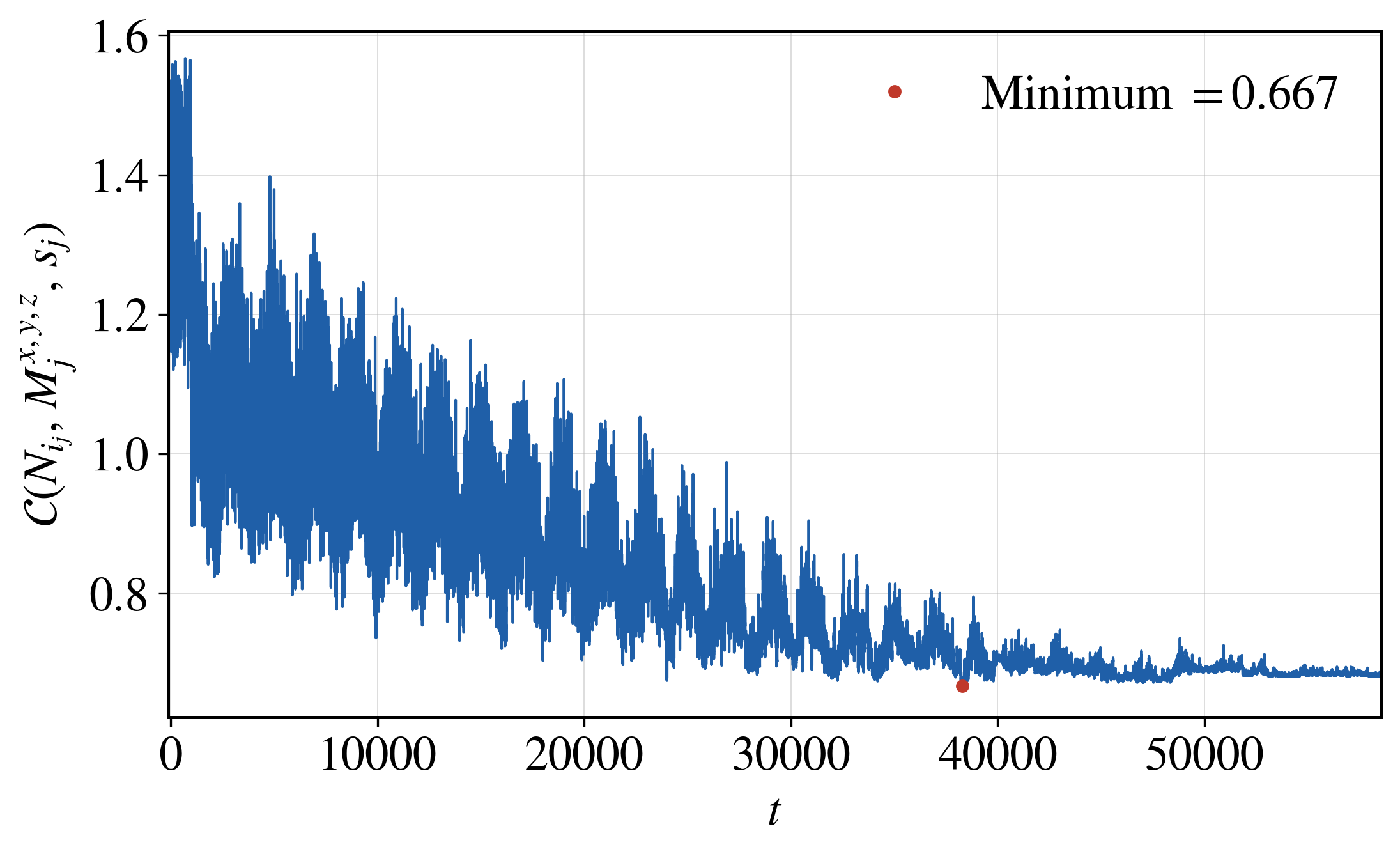}
  \caption{Cost function $\mathcal{C}$ as a function of the number 
    of proposed configurations $t$ during the 
    simulated annealing optimization. The red dot indicates the minimal
  cost found during the search.}
  \label{fig:annealing}
\end{figure}

\subsection{Optimal Bonsai string association}
\label{Bonsai optimization}

Beyond the choice of fermion-to-qubit mapping, the specific assignment of 
Majorana strings to fermionic lattice sites has a significant impact on the 
resulting circuit depth. Different assignments can lead to markedly different 
gate counts due to Pauli cancellations and routing overheads. To quantify 
these effects, we introduce a cost function that accounts for both Pauli 
weights and hardware-induced \textsc{swap} operations,
\begin{align}
\label{cost}
\mathcal{C}&= \mathcal C_0\sum_{j}  h_j C_j ,
\\  C_j &= \sum_{i_j}  2 (N_{i_j}-1)    
 + \alpha_1\,M^{x,y}_j + \alpha_2\,M^z_j + \alpha_3\,s_{j}, \nonumber
\end{align}
where $N_{i_j}$ denotes the number of nontrivial Pauli operators in the $i$-th 
string associated to the coupling $h_j$, $M^{x,y}_j$ is the total number of $X$ and $Y$ gates , $M^z_j$ the total number of $Z$ gates, and $s_j$ is the number of required \textsc{swap} operations for coupling $h_j$ (in our case, $U$ and $J$). 
The global coefficient $\mathcal C_0$ denotes the average $\mathrm{CZ}$ gate error observed on the \texttt{ibm\_boston} backend at the time of writing. The other coefficients $\alpha_1 = 0.24 $, $\alpha_2 = 0.48$, $\alpha_3 = 6.71$ are set according to the noise associated with each native gate, normalized with  respect to $\mathcal C_0$. 

Since an exhaustive optimization over all possible string assignments is 
computationally infeasible, even for the relatively small lattice sizes 
considered here, due to the combinatorial growth of the configuration space, 
we employ a heuristic strategy based on simulated annealing 
\cite{annealing,annealing1,annealing2,annealing3, Yu_2025}. Starting from an 
arbitrary initial assignment, new configurations are proposed by randomly 
swapping pairs of Majorana strings. Moves that lower the cost function are 
always accepted, while higher-cost configurations are accepted with a 
probability governed by an annealing schedule $\beta(t)$, with 
$t \in [0, N_\mathrm{sweep}]$, where $N_\mathrm{sweep}$ denotes the total 
number of proposed configurations. 

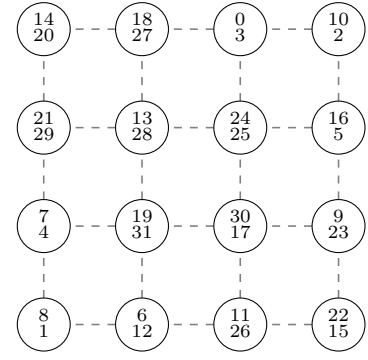
\begin{figure}[t]
  \centering
  \begin{tikzpicture}[
    scale=1.3,
    every node/.style={font=\scriptsize},
    dot/.style={circle, draw=black, fill=white, minimum size=7mm, inner sep=0pt},
    conn/.style={gray, dashed, line width=0.6pt}
    ]
    \def\listaA{{14,18,0,10,21,13,24,16,7,19,30,9,8,6,11,22}}
    \def\listaB{{20,27,3,2,29,28,25,5,4,31,17,23,1,12,26,15}}
    \foreach \i in {0,...,15}{
      \pgfmathtruncatemacro{\x}{mod(\i,4)}
      \pgfmathtruncatemacro{\y}{3 - int(\i/4)}
      \ifnum\x<3 \draw[conn] (\x,\y) -- (\x+1,\y); \fi
        \ifnum\y>0 \draw[conn] (\x,\y) -- (\x,\y-1); \fi
        }
        \foreach \i in {0,...,15}{
          \pgfmathtruncatemacro{\x}{mod(\i,4)}
          \pgfmathtruncatemacro{\y}{3 - int(\i/4)}
          \pgfmathparse{\listaA[\i]} \let\a\pgfmathresult
          \pgfmathparse{\listaB[\i]} \let\b\pgfmathresult
          \node[dot] at (\x,\y) {%
              \begin{tabular}{@{}c@{}}
                \a \\[-2pt] \b
              \end{tabular}
            };
        }
      \end{tikzpicture}
      \caption{Optimal Majorana string assignment on the $4\times 4$ 
        Hubbard lattice. Each node displays the indices $i$ and $j$ of the two 
        Majorana string pairs $(m_{2i}, m_{2i+1})$ and $(m_{2j}, m_{2j+1})$ associated with the 
      corresponding fermionic site.}
      \label{fig:bonsai_assignment}
    \end{figure}

The result of the optimization is shown in Fig.~\ref{fig:annealing} for the $4 \times 4$ lattice, while in Fig.~\ref{fig:bonsai_assignment} is presented the string-to-nodes association obtained from the optimization procedure. Compared to the naive index assignment, the optimized configuration reduces the cost function~(\ref{cost}) from $1.166$ to $0.667$, corresponding to roughly a $43\%$ decrease. Since the cost function is proportional to the total linear gate error accumulated over the full Hamiltonian evolution, this translates directly into a significant suppression of circuit noise. We note, however, that the actual gate fidelity depends exponentially on the per-gate error rate, and that our estimate assumes a uniform error rate across all qubits; the effective reduction in infidelity may therefore differ from the linear estimate provided by the cost function.
The full algorithmic details and the 
parameter choices are reported in Appendix \ref{app:dimostrazione}.

Unlike the approach of Ref.~\cite{Yu_2025}, which explores the full space of 
Clifford-equivalent mappings, our optimization is restricted to permutations 
of Majorana string assignments within the fixed Bonsai tree structure. 
Consequently, the tree structure of the mapping is naturally preserved: we 
simply associate the physical sites to different pairs of strings generated by the same Bonsai 
tree, so that only the explicit form of the Hamiltonian changes, while the 
mapping topology remains unaffected. Once routing costs are included in the 
optimization, Algorithm~\ref{alg:simulated_annealing} becomes highly 
efficient at runtime, since it starts from configurations in which overhead 
operations are already minimized and Steiner tree identification is a 
well-conditioned problem, as confirmed by our numerical results in Section \ref{section:classical_results}.

While the Clifford-based approach of~\cite{Yu_2025} is in principle more 
general, as it searches for optimal mappings beyond the ternary tree 
structure, it introduces practical challenges that become increasingly 
severe as the system size grows. First, the application of general Clifford 
operations may yield fermionic operators with highly non-local support, 
making the identification of efficient Steiner tree approximations 
progressively harder for larger lattices. Second, and more critically for 
near-term hardware, the Clifford circuit used to reach the optimal mapping 
representation must also be applied to prepare the initial fermionic state. 
We further note that the depth of this state preparation circuit is not 
analyzed in~\cite{Yu_2025}, and may scale non-trivially with system size. 
Crucially, this overhead is paid entirely during state preparation, before 
time evolution begins, and could significantly compromise the fidelity of 
the simulation on near-term hardware. 

By contrast, within the Bonsai 
framework, fermionic Fock states remain in the computational basis by 
construction~\cite{Miller}, so no additional operations are required once 
an optimal string assignment is found.

 \begin{figure*}[t]
    \centering
    \includegraphics[width=2.0\columnwidth]{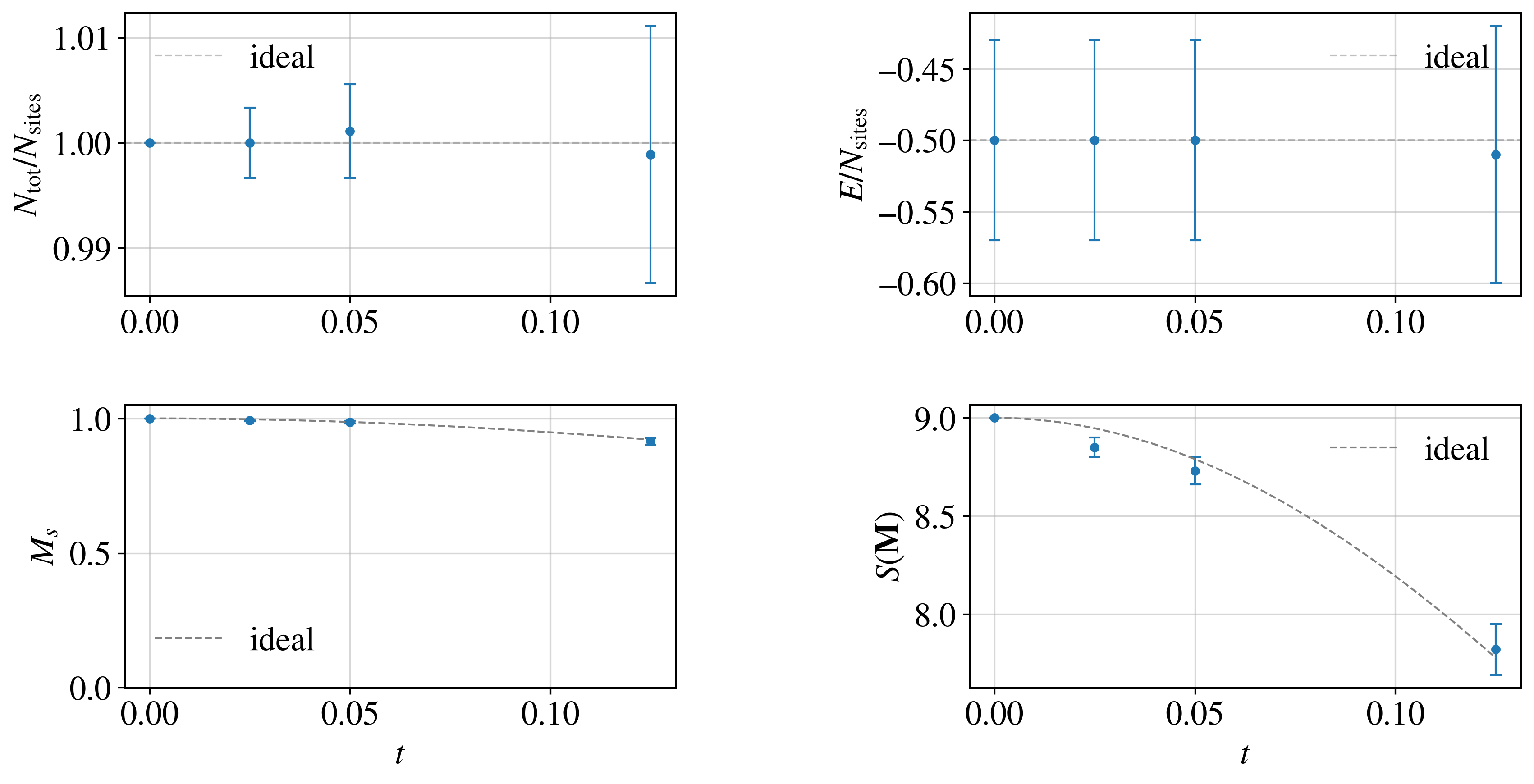}
    \caption{Global observables, Eqs.~\eqref{eq:energy}--\eqref{eq:structure}, as a function of evolution time for the ideal $3 \times 3$ lattice, implemented with our optimal quantum algorithm, compared against the classical TeNPy benchmark (dashed lines). The two methods exhibit an excellent agreement in all the observables at every time, except of the structure factor at $t=0.025$, which is however consistent with the tensor network approach within $2\sigma$. Clearly, this effect is due to the approximation performed by the qDRIFT evolution: once the qDRIFT precision is doubled, even this value approach the classical value in $1 \sigma$.}
    \label{fig:3x3_benchmark}
\end{figure*}

\subsection{Preliminary measurements and QEM}

In our work, we consider the strongly correlated regime, setting $J = 0.5$ 
and $U = 2$, corresponding to $U/J = 4$.
The choice of the parameters is determined by the limited time simulation on the real QPU, especially if we consider large systems as a $6 \times 6$ lattice. In particular, if we want to simulate the Néel state and magnetization decays, it is crucial to select specific parameters in order to remain in the strongly correlated regime, but, at the same time, the kinetic term must become competitive with respect to the on-site interaction to speed up the dynamics before the hardware noise intervenes. 

To verify the correctness of the quantum code, we time-evolved the vacuum 
state for $t = 0.125$  with $\epsilon_\mathrm{qDRIFT}= 4\cdot 10^{-3}$ through a $3 \times 3$ 
(spinful) Hubbard Hamiltonian with only on-site interaction on a classical laptop.
Importantly, notice that  times are expressed in units of the inverse of the hopping term $1/J$.

The state was confirmed to remain unchanged, with the expectation values of all local occupation numbers registering an empty site after the simulation, as expected.

We then prepared and time-evolved an initial Néel state with the full Hamiltonian. Since the Néel state is an antiferromagnetic configuration close to the exact ground state of the Hubbard Hamiltonian, we expect perturbations due to the hopping term in the local occupation numbers and magnetizations. These observables were measured 
at evolution times $t = 0, 0.025, 0.05, 0.125$  with $n_{\mathrm{shots}} = 1000$ 
repetitions per circuit. Increasing $n_{\mathrm{shots}}$ or the evolution 
time further was not feasible at this stage due to computational constraints.

The numerical results are reported in Appendix \ref{app:sim_results}, where it is possible to check the validity of the designed algorithm in detail.

In this section, we simply show how our simulation performs for the four global observables defined below:

\begin{enumerate}

    \item Energy per site
        \begin{equation}
            \frac{E}{N_{sites}}= \frac{\langle\hat{H}\rangle}{N_{sites}},
            \label{eq:energy}
        \end{equation}
        
        where $\hat{H}$ is the Hamiltonian defined in Eq.~\ref{eq:hamiltonian}.
    
    \item Occupation per site 
        \begin{equation}
            \frac{N_{tot}}{N_{sites}}= \frac{\langle\hat{N}\rangle}{N_{sites}},
        \end{equation} 
        
        where $\hat{N}$ is the operator of the total number of particles.
    
    \item Staggered magnetization
        \begin{equation}
            M_s = \frac{1}{L^2} \sum_{i} (-1)^{r_{i,x}+r_{i,y}} \braket{S^z_i}.
        \end{equation}
    
    \item Structure factor
        \begin{equation}
            S(\boldsymbol{q}) = \frac{1}{L^2} \sum_{i,j}
            e^{i \boldsymbol{q} \cdot (\boldsymbol{r}_i - \boldsymbol{r}_j)}
            \langle S^z_i S^z_j \rangle,
            \label{eq:structure}
        \end{equation}
        over the independent momenta $\boldsymbol{q}$ of the first Brillouin zone, where $\boldsymbol{r}_i$ denotes the real-space position of site $i$.  
\end{enumerate}

In particular, in the Brillouin zone, there exist 3 highly symmetrical points, the $\Gamma = (0,0)$, $X = (\pi, 0)$, and $M = (\pi, \pi)$ points, that determine the particular configuration in which the system sits. 
In our case of interest, if the $M$ point provide a peak in the structure factor, the system is in an antiferromagnetic order.
This follows directly from the structure of the correlations: at $\boldsymbol{q} = M$ the phase factor $e^{i\boldsymbol{q}\cdot(\boldsymbol{r}_i-\boldsymbol{r}_j)} = (-1)^{(x_i-x_j)+(y_i-y_j)}$ exactly compensates the alternating signs of $\langle S^z_i S^z_j \rangle$, yielding constructive summation at $M$. By contrast, at $\Gamma$ the phase factor is unity and the antiferromagnetic correlations cancel pairwise; an analogous cancellation occurs at $X$.

By measuring these observables through the algorithm of our work on the $3 \times 3$ lattice, we can test the reliability of our simulation by comparing the results to the observables provided by TeNPy \cite{SciPostPhysCodeb.41}, a classical approach based on tensor networks. The comparison is reported in Fig.~\ref{fig:3x3_benchmark}. We notice that all the data are in excellent agreement with the classical prediction. The only remarkable measurement with a slight deviation is at $t=0.025$ in the structure factor, but is simply generated by the qDRIFT approximation. As a matter of fact, simply doubling the precision in the evolution time allows a better convergence from $2 \sigma$ to $1 \sigma$ for this last point. 

Regarding the qDRIFT error, a constant error bound was imposed for each 
time evolution step. Notably, all the global observables are numerically consistent with the theoretical predictions at all times, even without 
saturating the bound in Eq.~(\ref{eq:bound}). This is consistent with the 
fact that the bound refers to worst-case scenarios. Generally, the difference between the exact evolution and the qDRIFT decomposition depends on the commutator of the Hamiltonian and the density matrix of the initial state \cite{Campbell_2019}. In our case, the density matrix is a pure state created by applying number operators on the density matrix of the vacuum, and the number operators present in the Hamiltonian (see Sec. \ref{sec:hamiltonian}) commute with the majority of the terms, which positively affects the precision of the qDRIFT algorithm. 
As a consequence, for simulations on the QPU targeting a given precision in the 
measured observables, it is reasonable to employ fewer qDRIFT steps than 
those imposed by Eq.~(\ref{eq:bound}), while still obtaining a reliable 
approximation of the time-evolved state.

Having established the correctness of the qDRIFT protocol and the Bonsai mapping in an ideal setting, the primary challenge for real QPU execution remains addressing residual hardware noise. To this end, we integrated three well-known and general Quantum Error Mitigation (QEM) techniques: TREX (Twirled Readout Error Extinction), dynamical decoupling and ZNE (Zero-Noise Extrapolation) -- see also \cite{zimboras2025myths} for comparisons. The first two are both implemented via native Qiskit routines, for the latter we opted for a manual implementation of ZNE.
Specifically, we employed an optimized Richardson extrapolation, as suggested in \cite{PhysRevA.106.062436, cai2023practicalframeworkquantumerror, hoel1964optimal}, and reported in Appendix \ref{app:richardson}. For every time evolution, we implemented a combination of these techniques in order to minimize hardware noise.

\section{Results on quantum hardware}
\label{sec:results}

\subsection{Hardware implementation and circuit characterization}

The quantum simulation was executed on \texttt{ibm\_boston}, selected among the available backends for its superior single-qubit coherence times and gate fidelities, at the time of writing. Physical qubits were chosen to maximize gate fidelity over the active computational subspace. We simulate both the $4 \times 4$ and $6 \times 6$ Hubbard lattices, comparing the results to assess scalability-induced deviations in both physical observables and hardware performance, following the approach of Ref.~\cite{alam2025programmabledigitalquantumsimulation}.

Both lattices were simulated using a qDRIFT error $\epsilon_\mathrm{qDRIFT} = 2 \cdot 10^{-2}$, except at $t = 0.125$, where the resulting gate count exceeded the practical circuit depth limit and a relaxed precision $\epsilon_\mathrm{qDRIFT} = 1 \cdot 10^{-1}$ was adopted. Each time configuration was sampled using $5000$ effective shots (i.e.~to obtain a statistical error equivalent to $5000$ shots, but in practice ZNE required from $10^4$ to $10^5$ shots, depending on the form of the extrapolation involved) for the $4 \times 4$ lattice and $1000$ effective shots for the $6 \times 6$ lattice. A linear ZNE extrapolation was employed at $t = 0.025$, while a quadratic extrapolation was used at $t = 0.05$ and $t = 0.125$, where deeper circuits exhibit a non-linear noise scaling with circuit depth. As noted in Ref.~\cite{PhysRevA.106.062436}, the polynomial order of the ZNE extrapolation should remain moderate to avoid excessive shot overhead, a constraint that becomes particularly stringent at the current stage of quantum hardware development.

The hardware resource budget for the $t = 0.125$ configuration of the $6 \times 6$ lattice is summarized in Table~\ref{tab:gate_count}, reporting gate counts and estimated execution times for the ten most heavily utilized qubits.

\begin{table}[t]
\centering
\begin{tabular}{cccccccc}
\toprule
Qubit & SX & $R_X$ & CZ & $R_{ZZ}$ & $X$ & Total & $T_{\mathrm{ex}}$ ($\mu$s) \\
\midrule
91  & 111 & 38 & 241 & 4  & 4 & 398 & 20.07 \\
111 & 105 & 51 & 230 & 7  & 0 & 393 & 19.00 \\
89  &  85 & 37 & 195 & 4  & 0 & 321 & 15.98 \\
113 &  81 & 41 & 160 & 21 & 6 & 309 & 13.66 \\
98  &  88 & 14 & 186 & 2  & 2 & 292 & 15.53 \\
112 &  80 & 12 & 168 & 5  & 0 & 265 & 13.98 \\
69  &  70 & 16 & 146 & 5  & 4 & 241 & 12.30 \\
93  &  63 & 20 & 139 & 6  & 2 & 230 & 11.53 \\
92  &  59 & 14 & 144 & 3  & 0 & 220 & 11.68 \\
90  &  62 &  6 & 143 & 2  & 0 & 213 & 11.71 \\
\midrule
$T_{\mathrm{circ}}$ ($\mu$s) & & & & 130.36 \\
\bottomrule
\end{tabular}
\caption{Gate count and circuit execution time $T_{\mathrm{ex}}$ (the active
execution time on a qubit) for the 10 most active
physical qubits in the $6\times6$ Hubbard model simulation
($t=0.125\,J^{-1}$). Only error-accumulating native gates are reported:
$Sx$, $R_x$, $CZ$, $R_{zz}$, and $X$. Virtual $R_z$ gates are omitted, as
they are implemented in software and contribute no gate error. We note
that $CZ$ and $R_{zz}$ gates are shared between two qubits, so the same
two-qubit gate may be counted twice in the gate count (once for the
control qubit and once for the target). $T_{\mathrm{ex}}$ is computed as the sum of the durations of every gate applied to that qubit; $T_{\mathrm{circ}}$ is the total duration of the quantum circuit. The large difference between the two quantities arises from idle time, i.e.\ the intervals of inactivity on a given qubit between successive operations. In any case, $T_{\mathrm{circ}}$ remains below the average hardware decoherence time $T_2\approx340\,\mu s$, but can contribute in a non-negligible way to the total gate infidelity.
However, this type of decoherence during idle intervals is further mitigated by
dynamical decoupling, which is particularly well suited to circuits with
long periods of inactivity.}
         
\label{tab:gate_count}
\end{table}

\subsection{Global observables}

As operated in the previous sections, in this section we only discuss the global observables measured for the four times considered ($t=0,0.025,0.05,0.125$). For further details about local quantities, see Appendix \ref{app:sim_results}.

\begin{figure*}[t]
    \centering
    \includegraphics[width=2.0\columnwidth]{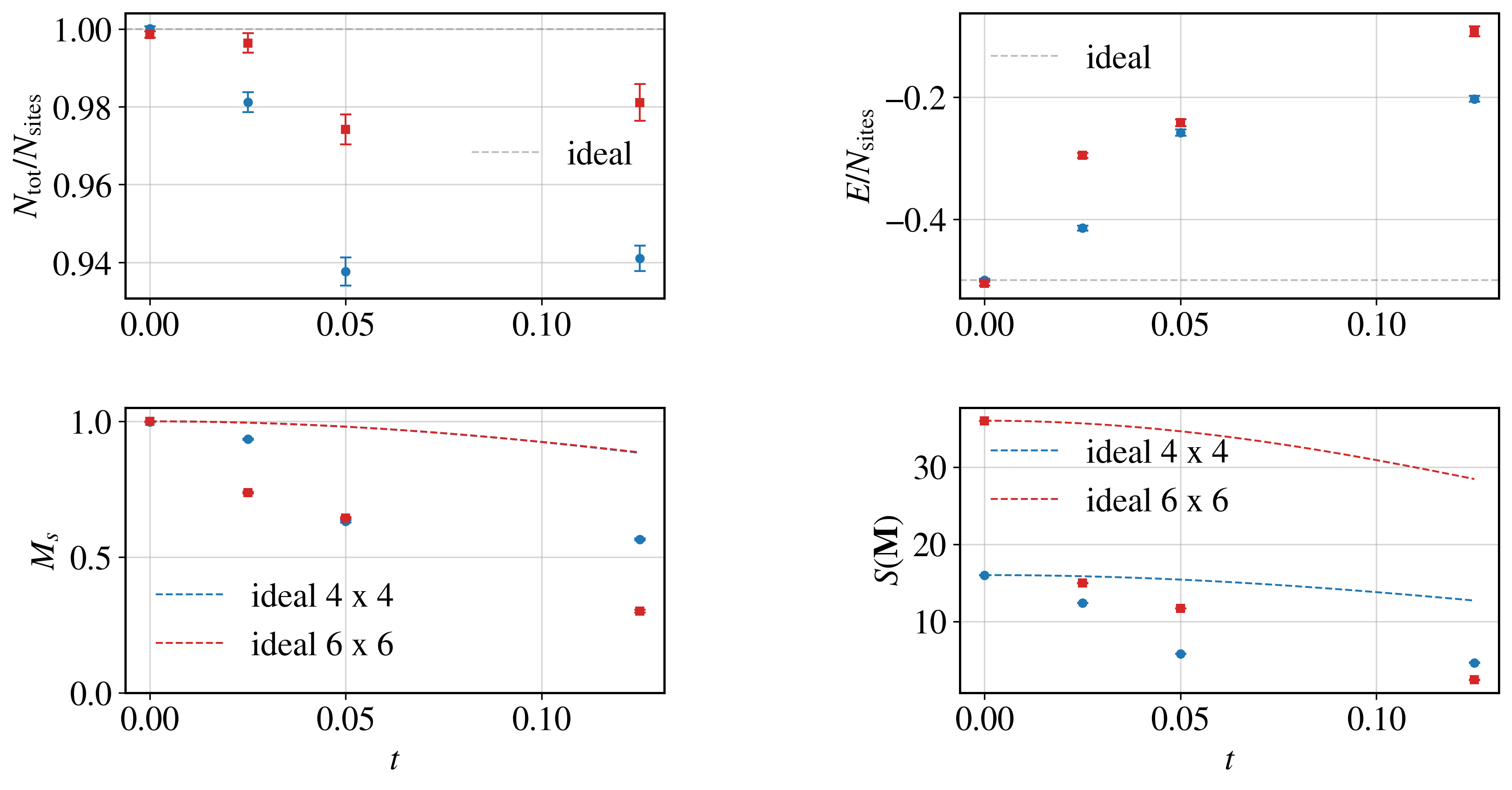}
    \caption{Global observables as a function of evolution time for the $4 \times 4$ and $6 \times 6$
    lattices, compared against the classical TeNPy benchmark (dashed lines). The $4 \times 4$ results
    remain closer to the ideal values throughout, though noticeable deviations emerge at later times
    in both cases. The apparent recovery of the occupation number toward the half-filling value
    $N_\mathrm{tot}/N_\mathrm{sites} = 1$ at late times is an artifact of hardware noise: in the
    fully depolarized limit, each spin species independently saturates to
    $\langle n_{\uparrow,\downarrow}\rangle \to 1/2$, yielding $N_\mathrm{tot}/N_{sites} \to 1$
    regardless of the physical state.}
    \label{fig:global_observables}
\end{figure*}
The deviation of global observables from the classical TeNPy benchmark~\cite{SciPostPhysCodeb.41} for both lattices is shown as a function of time in Fig.~\ref{fig:global_observables}. At late times, the apparent recovery of the occupation number toward the half-filling value $N_\mathrm{tot}/N_\mathrm{sites} = 1$ of the $6 \times 6$ lattice is not a physical effect: in the fully depolarized limit, each spin species independently saturates to $\langle n_{\uparrow,\downarrow}\rangle \to 1/2$, yielding $N_\mathrm{tot}/N_{sites} \to 1$ regardless of the physical state. Once this saturation regime is entered, the simulation no longer carries meaningful information, and we therefore stop the time evolution at $t = 0.125$.
Three primary mechanisms drive the observable degradation. The first is
decoherence: as reported in Tab.~\ref{tab:gate_count}, the total duration
of the circuit, $T_{\mathrm{circ}}$, is $130.36\,\mu s$. Compared with the
average decoherence time measured on the chip, $T_2 = 340\,\mu s$, this
duration alone accounts for a non-negligible loss of fidelity on the final
state,
\begin{equation}
    F_{T_2} \approx e^{-T_{\mathrm{circ}}/T_2} \approx 0.68.
    \label{eq:decoherence_fidelity}
\end{equation}
However, Tab.~\ref{tab:gate_count} also shows that $T_{ex}$, the time
during which the QPU is actually active on the qubits, is only a fraction of $T_{\mathrm{circ}}$; for most of the circuit duration, qubits remain idle on the chip. The dynamical decoupling technique introduced in the previous section is particularly well suited to circuits with long idle periods, and plays a fundamental role in mitigating decoherence on otherwise-inactive qubits during these windows. In conclusion, the DD-mitigated fidelity loss is therefore expected to be substantially lower than the bare bound in Eq. \ref{eq:decoherence_fidelity}. Since the case reported in Tab.~\ref{tab:gate_count} represents the worst-case scenario in this work, and the circuits studied elsewhere in the paper generally exhibit durations well below the one highlighted here, decoherence remains a well-confined effect throughout our simulations.

The second aspect to be considered is the accumulation of gate infidelity: for qubit 91 of the $6 \times 6$ lattice at $t = 0.125$ (see Table~\ref{tab:gate_count}), the total circuit fidelity estimated via Eq.~\ref{eq: gate_fidelity} is approximately $0.72$, with decoherence subdominant at this depth relative to the device coherence times. Since local magnetization observables involve joint readout over $3$--$5$ qubits, the effective fidelity is correspondingly lower, explaining the rapid signal deterioration at late times. The third mechanism is the limited effectiveness of ZNE on real hardware~\cite{PhysRevA.106.062436}: while ZNE performs well on \texttt{Qiskit AerSimulator} noise models, physical devices may exhibit non-Markovian or hardware-specific noise processes not captured by standard error models, reducing the extrapolation accuracy.
Notwithstanding these limitations, the antiferromagnetic structure remains qualitatively preserved at all simulated times, motivating the analysis of the spin structure factor.

\subsection{Spin structure factor}

\begin{figure*}[t]
    \centering

        \includegraphics[width=2.0\columnwidth]{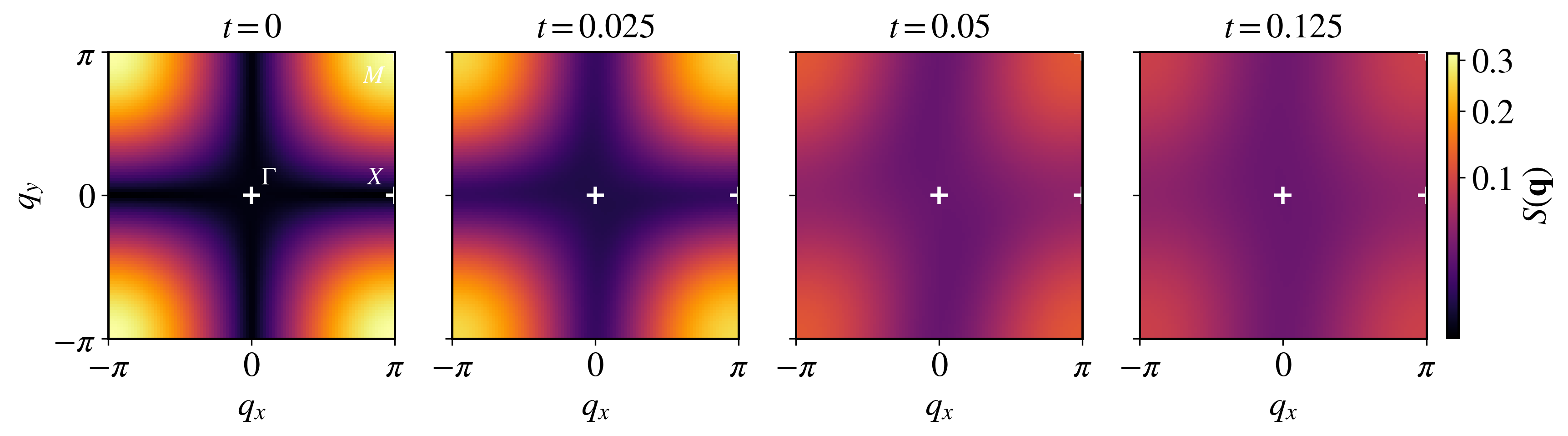}  
        
    \caption{Spin structure factor $S(\boldsymbol{q})$ over the first Brillouin zone of the square lattice at times
    $t=0$, $t=0.05$, $t=0.10$, and $t=0.125$, for the $4\times 4$ lattice. The values are reported with a square root  scale.
    The peak at the $M$ point persists throughout the entire evolution, reflecting the robustness of the
    antiferromagnetic ordering wavevector: the phase factor in Eq.~\ref{eq:structure} evaluated at
    $\boldsymbol{q} = M$ exactly compensates the alternating signs of the spin correlations, producing
    constructive interference at $M$ and destructive cancellation at $\Gamma$ and $X$.}
    \label{fig:struttura_4x4}
\end{figure*}

To characterize the antiferromagnetic order more systematically, we compute the spin structure factor defined in Eq.~\ref{eq:structure}.
Since the spin correlations $\langle S^z_i S^z_j \rangle$ are evaluated on finite lattices, a direct discrete Fourier transform introduces spurious Gibbs oscillations arising from the sharp truncation of the correlation function at the boundary. To suppress these artifacts while preserving the physical signal, we apply a two-dimensional Hann window to the real-space correlation matrix prior to the Fourier transform, following standard practice in finite-size spectral analysis. The antiferromagnetic peak at $M$ survives the windowing procedure with its position and sign unchanged, acquiring only a modest broadening due to the finite support of the Hann kernel.

\begin{figure*}[t]
    \centering
    \includegraphics[width=2.0\columnwidth]{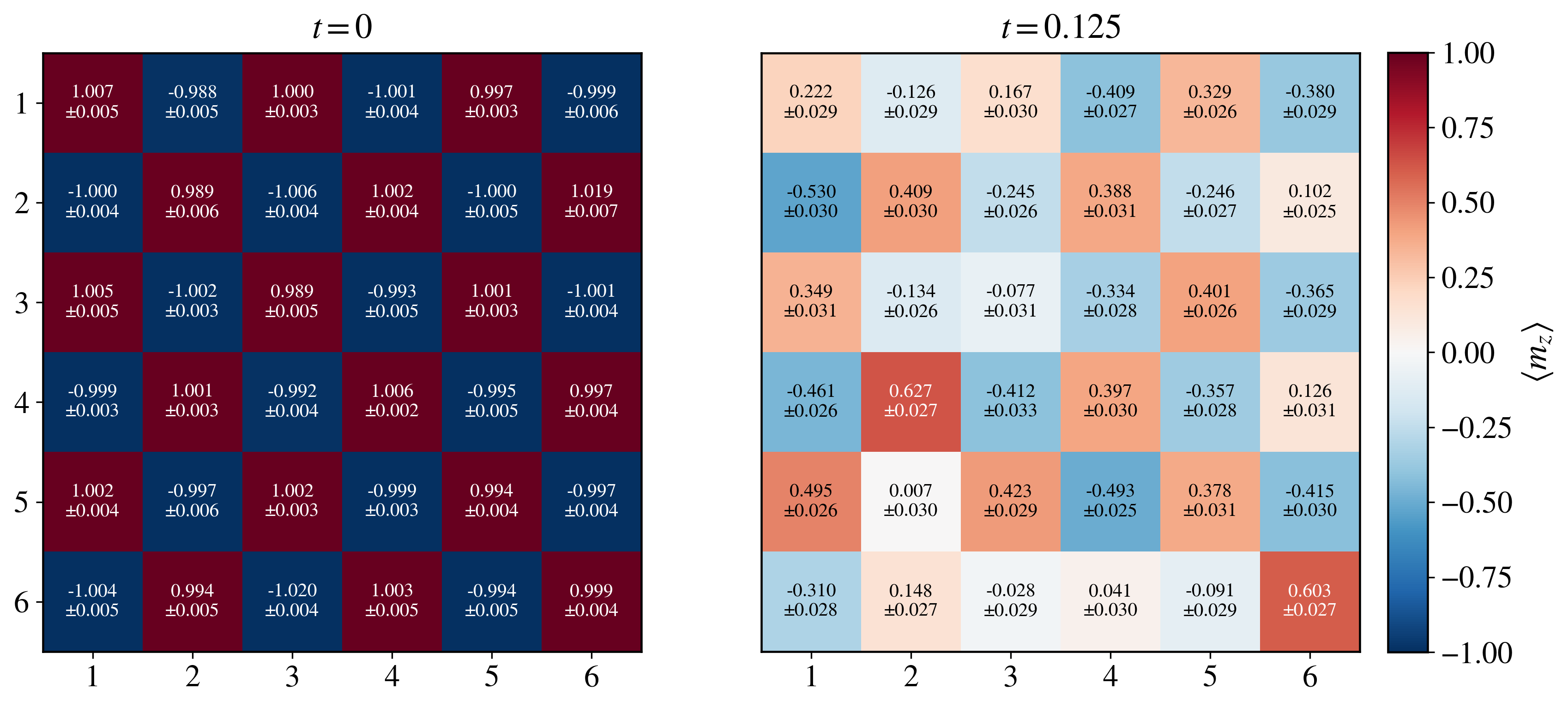}
    \includegraphics[width=2.0\columnwidth]{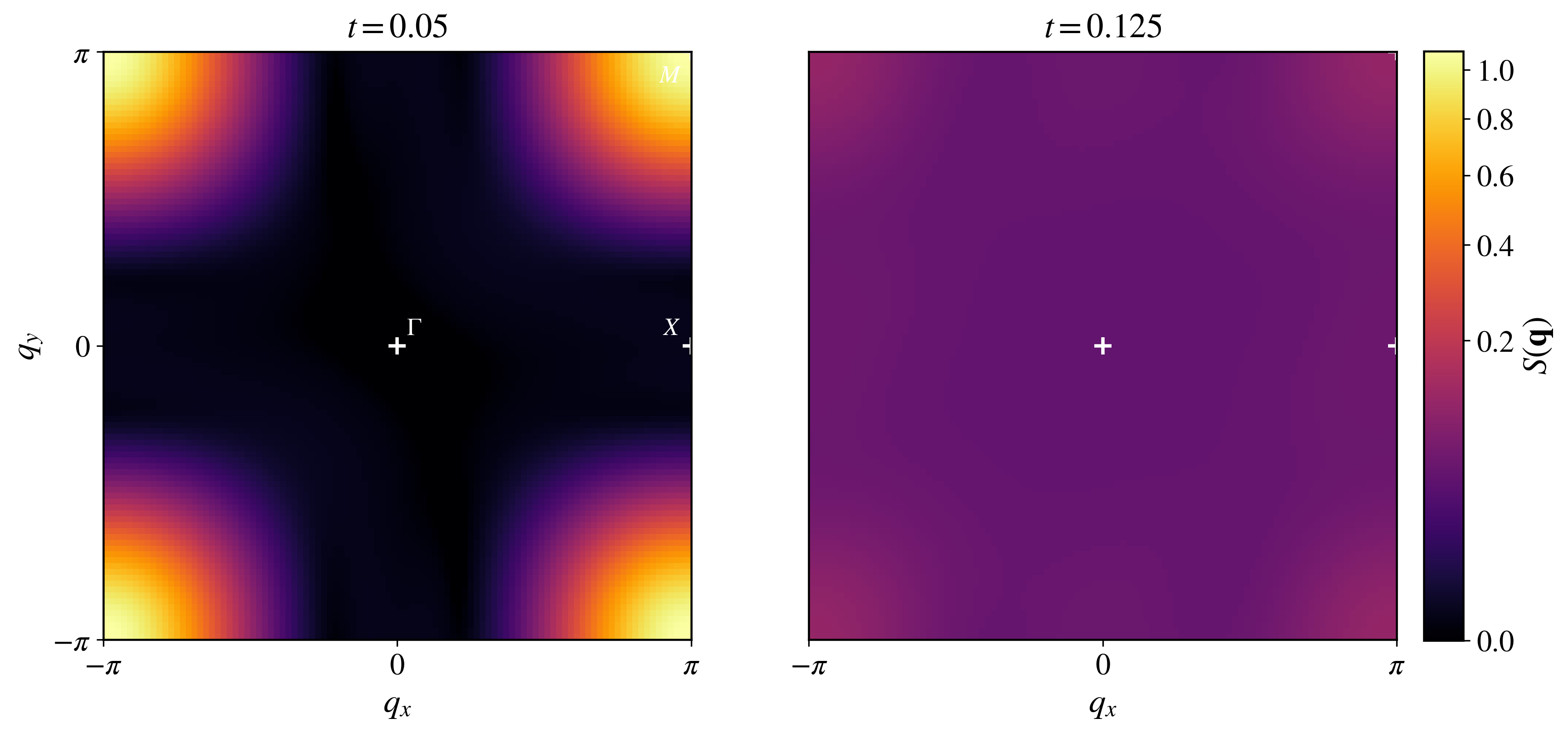}
    \caption{Local magnetization on the $6 \times 6$ square lattice at $t = 0$  and $t = 0.125$ (top figure),
    and the corresponding spin structure factor over the first Brillouin zone (bottom figure), reported in a power scale with an exponent of $0.4$. The initial
    N\'{e}el state is correctly prepared via the Bonsai mapping of Ref.~\cite{Miller} and subsequently
    evolved under the qDRIFT protocol. Despite gate infidelities introducing quantitative deviations
    from both the classical TeNPy simulation and the results of Ref.~\cite{alam2025programmabledigitalquantumsimulation},
    the antiferromagnetic structure is qualitatively preserved throughout the time evolution.}
    \label{fig:lattice_6x6}
\end{figure*}

The high-symmetry points $\Gamma = (0,0)$, $X = (\pi, 0)$, and $M = (\pi, \pi)$ are marked in Figs.~\ref{fig:struttura_4x4} and~\ref{fig:lattice_6x6} with white crosses. At $t = 0$, $S(\boldsymbol{q})$ exhibits a sharp peak at $M$, consistent with the antiferromagnetic ordering wavevector of the N\'{e}el state.
In particular, we notice a discrepancy  in the values of the $M$ point of Figs.~\ref{fig:struttura_4x4} and  \ref{fig:global_observables} (see also \ref{fig:magnetizzazioni}).  However, we can check easily that the Hann windowing is responsible of this behaviour. If we consider the initial Neél state for the $4 \times 4$ lattice, we compute directly the $S_{win}$ mitigated value for the $M$ point:
\begin{equation}
    S_{win}(\boldsymbol{M}) = \frac{( \sum^{N_{sites}}_i W(i) )^2}{L^2} = \frac{(S_x S_y)^2}{L^2}= 0.31,
\end{equation}
where $W(i) = S_x S_y$, and $S_x = S_y = \sum^{L}_n  \sin^2({\frac{n \pi}{L -1}})$.

This is exactly what is reproduced in Fig.~\ref{fig:struttura_4x4} at $t=0$.
 
The time evolution of $S(\boldsymbol{M},t)$, shown in Fig.~\ref{fig:global_observables}, reveals a monotonic decrease of the peak amplitude. This decay is qualitatively consistent with the erosion of antiferromagnetic order driven by the hopping term; however, at the accessible evolution times, hardware-induced demagnetization contributes a systematic bias of the same sign and cannot be disentangled from the physical signal. As expected, the largest deviations from the classical TeNPy benchmark are observed for the $6 \times 6$ lattice, which exhibits a steep drop in $S(\boldsymbol{M},t)$ already at early times, inconsistent with the mild decrease predicted classically. 

We furthermore declare that we do not report per-point uncertainties on $S(\boldsymbol{q})$ across the full Brillouin zone. Standard error propagation is not applicable here, as the correlators $\langle S^z_i S^z_j \rangle$ entering the sum are statistically correlated, and these inter-correlator covariances are non-negligible near high-symmetry points where contributions interfere constructively. This is confirmed by the fact that in data of Fig.~\ref{fig:3x3_benchmark} we directly computed the value of $S(\boldsymbol{M})$, taking into account the correlations. Even if the expected value remains  substantially unchanged, the variance increased by $10$ times with respect to the case in which we consider the spatial correlations as independent variables.

\subsection{Analysis on simulation limitations}

We identify three systematic limitations of the present implementation. First, while qDRIFT is the preferred decomposition strategy for higher-dimensional systems with long-range interactions~\cite{Campbell_2019}, the partial gate cancellations enabled by second-order Trotter--Suzuki decompositions and $f$-SWAP networks~\cite{alam2025programmabledigitalquantumsimulation} are not available here, as qDRIFT does not admit the merging of consecutive gates in the same way as deterministic product formulas. Nevertheless, when the Hamiltonian contains a large number of commuting terms --- as in the Hubbard model --- pairs of Pauli rotations sharing the same Majorana string can still be grouped and merged into a single rotation, providing partial gate count reduction even within the stochastic decomposition framework.

Second, the statistical budget of at most $10^5$ shots per configuration constrains the achievable mitigation precision, since ZNE performance scales favorably only with significantly larger shot counts.

Lastly, we stress that the primary contribution of this work is an efficient fermionic encoding of the 2D Hubbard model on heavy-hexagon hardware, rather than the development of advanced quantum error mitigation pipelines. In this context, the residual discrepancy is a quantifiable outcome of the current mitigation strategy, not an uncontrolled systematic. Indeed, recent work~\cite{alam2025programmabledigitalquantumsimulation} has demonstrated faithful signal recovery for circuits containing up to 4372 two-qubit gates --- nearly double the depth of our largest circuit (2775 two-qubit gates, see Tab.~\ref{tab:gate_count}) --- through a combination of advanced techniques,
while also finding that ZNE underperforms relative to these methods at comparable circuit depths. Applying such techniques to the present simulation represents a natural and well-motivated extension of this work.

However, regardless of these quantitative limitations, the persistence of the peak at $M$ across all simulated times confirms that the antiferromagnetic ordering wavevector is robustly encoded in the quantum state throughout the evolution.


\section{Conclusions}
\label{sec:conclusions}

The overarching goal of quantum computing is to execute algorithms that
outperform their classical counterparts on tasks of practical relevance,
with the simulation of correlated fermionic systems being among the most
compelling near-term targets. In this work, we addressed this challenge in
the context of the two-dimensional Hubbard model, focusing on the design
and optimization of fermion-to-qubit mappings tailored to the heavy-hexagon
topology of IBM superconducting quantum processors.

In the first part of this work, we benchmarked several fermion-to-qubit
transformations and demonstrated that the Bonsai mapping minimizes both the
Pauli weight and the \textsc{swap} overhead on heavy-hexagon hardware more
effectively than the Jordan--Wigner and Bravyi--Kitaev transformations.
Although the Bravyi--Kitaev mapping initially exhibits a lower Pauli weight,
this advantage is eliminated once routing costs are accounted for,
confirming the Bonsai mapping as the most hardware-efficient baseline
transformation for this architecture.

We further reduced the linear gate error cost by nearly $50\%$ through a
simulated annealing optimization that permutes the Bonsai Majorana string
assignments to simultaneously minimize the \textsc{swap} overhead and the
Pauli weight. While less general than the approach of Ref.~\cite{Yu_2025},
this algorithm provides an accurate estimate of the \textsc{swap} overhead
within a manageable runtime, without introducing any change of
representation in the fermionic Fock basis.

A natural extension of this work would be to combine our approach with that
of Ref.~\cite{Yu_2025} in a two-stage optimization. In the first stage, the
simulated annealing search is restricted to the Bonsai configuration space,
yielding a hardware-aware optimal assignment as the starting point. In the
second stage, Clifford transformations are applied to explore the mapping
space beyond the Bonsai structure, guided by a cost function that weights
each Hamiltonian term by its interaction strength $h_j$. This choice is
physically motivated by the qDRIFT sampling strategy, in which each term is
selected with probability proportional to $h_j$: reducing the gate cost of
the dominant terms therefore yields a disproportionately large reduction in
the total circuit noise accumulated during the simulation. Such a
hardware- and interaction-aware cost function could determine whether a
mapping more efficient than the Bonsai transformation exists for the
heavy-hexagon topology, while the state-preparation overhead would be
substantially reduced relative to a purely Clifford-based approach, since
the search begins from an already optimized Bonsai configuration~\cite{Miller}.

In the second part of this work, we performed real-device simulations of
the Hubbard model on the $4 \times 4$ and $6 \times 6$ lattices, which are
among the largest two-dimensional systems currently simulated on quantum
hardware, using standard quantum error mitigation techniques comprising
ZNE, TREX, and dynamical decoupling. Despite
the quantitative disagreement with the ideal evolution at late times ---
an expected consequence of the simplified mitigation pipeline adopted ---
the simulation preserves a physically meaningful signal up to
$t = 0.125\,J^{-1}$.
The antiferromagnetic ordering
wavevector is robustly identified throughout, confirming the viability of
the Bonsai encoding for large-scale two-dimensional simulations.

These results suggest several directions for future work. On the mitigation
side, the incorporation of more sophisticated techniques 
represents the most direct route to
extending the accessible evolution time. On the algorithmic side, using Quantum Signal Processing 
techniques \cite{low2017optimal,motlagh2024generalized} techniques or combining
the two-stage Clifford optimization described above with interaction-aware
qDRIFT sampling could further reduce the circuit depth, enabling longer
coherent evolution within the noise budget of current hardware. Together,
these improvements may allow the present framework to be extended to more
complex condensed matter systems and to dynamical regimes currently
inaccessible without advanced error mitigation.


\begin{acknowledgments}
This work is supported
by QUART\&T, a project funded by the Italian Institute of Nuclear Physics (INFN) within the Technological and Interdisciplinary Research Commission (CSN5) and Theoretical Physics Commission (CSN4), 
by the Italian National Quantum Science and Technology Institute through the PNRR MUR Project under Grant PE0000023-NQSTI and by the Italian Research Center on High Performance Computing, Big Data and Quantum Computing through the PNRR MUR Project under Grant  CN00000013-ICSC. We acknowledge the use of IBM Quantum services for this work. The views expressed are those of the authors, and do not reflect the official policy or position of IBM or the IBM Quantum team.
\end{acknowledgments}

\appendix

\section{IMPROVED TAKAHASHI-MATSUYAMA ALGORITHM}
\label{app:TM-alg}

In this appendix, we describe in detail the heuristic procedure used to estimate the \textsc{swap} overhead for the three fermion-to-qubit mappings of Tab.~\ref{tab:mappature}, summarized in Algorithm~\ref{alg:swap_count}. As outlined in Sec.~\ref{section:classical_results}, the algorithm proceeds in two stages.
In the first stage, contiguous subsets of qubits on which the Pauli string acts are grouped into blocks, within which no \textsc{swap} operations are required. Each block is then contracted into a single supernode, yielding a reduced graph that preserves the original hardware connectivity between blocks.
In the second stage, an approximate Steiner tree is constructed on the reduced graph. Starting from an arbitrary supernode, the algorithm iteratively connects the supernode whose topological distance from the current tree support is minimal, extending the tree at each step. Shortest paths between supernodes are computed via the multi-source Dijkstra algorithm~\cite{Dijkstra1959}. Once a supernode is incorporated into the tree, it is removed from the list of remaining terminals. The procedure terminates when all terminals have been connected, and the total \textsc{swap} cost is read off from the number of intermediate nodes traversed by the spanning tree.

\begin{algorithm*}[!ht]
\caption{Heuristic Takahashi-Matsuyama algorithm on Supergraph}
\label{alg:swap_count}
\KwIn{Operator qubit set $\mathcal{Q}$, hardware connectivity graph $\mathcal{G}$}
\KwOut{Number of required SWAPs $s$, number of terminal supernodes $n_b$}
\BlankLine

\tcp{Phase 1: Supergraph contraction}
Contract $\mathcal{G}$ into supergraph $\tilde{\mathcal{G}}$ by merging physically adjacent qubits in $\mathcal{Q}$ into supernodes\;
Extract terminal supernodes $\mathcal{T} = \{\tilde{v}_1, \ldots, \tilde{v}_{n_b}\}$ corresponding to the operator support\;
$n_b \leftarrow |\mathcal{T}|$\;
\BlankLine
\If{$n_b \leq 1$}{
    \Return $0$\;
}
\BlankLine

\tcp{Phase 2: Greedy Steiner tree approximation}
$\mathcal{S} \leftarrow \{\tilde{v}_1\}$ \tcp*{Steiner tree node set, seeded with first terminal}
$\mathcal{R} \leftarrow \{\tilde{v}_2, \ldots, \tilde{v}_{n_b}\}$ \tcp*{remaining terminals}
$\mathcal{F} \leftarrow \{\tilde{v}_1\}$ \tcp*{frontier: supernodes already connected}
\BlankLine
\While{$\mathcal{R} \neq \emptyset$}{
    $d^* \leftarrow \infty$,\quad $\pi^* \leftarrow \emptyset$,\quad $t^* \leftarrow \text{null}$\;
    \BlankLine
    \For{each $t \in \mathcal{R}$}{
        Compute shortest path via multi-source Dijkstra: $(d, \pi) \leftarrow \textsc{Dijkstra}(\tilde{\mathcal{G}},\, \mathcal{F},\, t)$\;
        \If{$d < d^*$}{
            $d^* \leftarrow d$,\quad $\pi^* \leftarrow \pi$,\quad $t^* \leftarrow t$\;
        }
    }
    \BlankLine
    \If{$t^* = \emph{null}$}{
        \textbf{break} \tcp*{remaining terminals unreachable}
    }
    \BlankLine
    \For{each node $v \in \pi^*$}{
        $\mathcal{S} \leftarrow \mathcal{S} \cup \{v\}$\;
        $\mathcal{F} \leftarrow \mathcal{F} \cup \{v\}$\;
    }
    $\mathcal{R} \leftarrow \mathcal{R} \setminus \{t^*\}$\;
}
\BlankLine

\tcp{Phase 3: SWAP count}
$\mathcal{P} \leftarrow \{ v \in \mathcal{S} \mid v \text{ is not a terminal supernode} \}$ \tcp*{Steiner points}
$s \leftarrow |\mathcal{P}|$\;
\BlankLine
\Return $s$\;
\end{algorithm*}

\section{qDRIFT DECOMPOSITION AND LIMITATIONS ABOUT THE TIME EVOLUTION}
\label{app:qDRIFT protocol}
In this Appendix, we briefly analyze the qDRIFT algorithm first proposed in \cite{Campbell_2019}, and we later discuss the parameters that must be monitored in the Hubbard numerical simulation.

The qDRIFT decomposition proceeds as follows. The Hamiltonian is first written as
\begin{equation}
\hat{H} = \sum_{j=1}^{L} h_j H_j,
\end{equation}
where the operators $H_j$ are Hermitian and normalized, and $h_j>0$. The target unitary is then approximated by
\begin{equation}
V_{\boldsymbol{j}} = \prod_{k=1}^{N} e^{-i \tau H_{j_k}},
\end{equation}
with $\tau \equiv t \lambda/N$ and $\lambda = \sum_j h_j $. Each $H_{j_k}$ is sampled independently with probability $p_j = h_j / \lambda$, and $\boldsymbol{j} = \{j_1,j_2 \dots, j_N\}$ represents the vector of the sampled terms. In the limit of large $N$, the approximation converges to the exact evolution operator.
with $\tau \equiv t \lambda/N$ and $\lambda = \sum_j h_j $. Each $H_{j_k}$ is sampled independently with probability $p_j = h_j / \lambda$, and $\boldsymbol{j} = \{j_1,j_2 \dots, j_N\}$ represents the vector of the sampled terms of the Hamiltonian. In the limit of large $N$, the approximation converges to the exact evolution operator.
In particular, the improved error bound of qDRIFT from \cite{david2025tightererrorboundsqdrift} 
\begin{equation}
    \epsilon_\mathrm{qDRIFT} \leq \frac{4 \lambda t^2}{N},
    \label{eq:bound}
\end{equation}
 is linear in the number of terms in the Hamiltonian, in contrast with Trotter--Suzuki decompositions, which require at least $\mathcal{O}(L^2)$ terms in the best cases. 

 
The accuracy of the approximation improves with increasing $N$, but practical implementations are constrained by hardware coherence times. In particular, the total circuit depth must remain below the dephasing time, which accounts for both energy relaxation and pure dephasing processes. For the \texttt{ibm\_boston} Heron r3 processor, the median dephasing time reported by IBM Quantum is $340~\mu\mathrm{s}$. Assuming an average single-gate duration of $100~\mathrm{ns}$, this sets an upper bound of approximately $10^{3}$ gates per qubit for the entire simulation.

Beyond this, a further limitation is represented by gate fidelity, which refers to how closely a real quantum gate approximates a theoretical one. The fidelity of a single gate is $F_{gate}=1-E_{gate}$, where $E_{gate}$ is the error incurred during the computation on a qubit by the gate. If we assume a depolarizing channel model, the total gate fidelity is given by:
\begin{equation}
    F_{tot}=\prod^n_{i=0}F_i,
    \label{eq: gate_fidelity}
\end{equation}
where $F_i$ is the fidelity of a single gate.
If we consider the simplest of the two models we want to simulate, the $4 \times 4$ Hubbard model, we can provide an estimate of the total number of qDRIFT steps we can implement. For 32 qubits, based on the results obtained from the fit, an average single excitation is given roughly by $15.5$ $Cz$ gates or $0.97$ $Cz$ per qubit at each step. If the average $Cz$ gate error that occurred in a circuit is $10^{-3}$, then for a gate fidelity no less than the $50\%$ we can implement approximately $N=600$ qDRIFT steps, which are equivalent to a total time of $0.25$  of simulation with a precision of $\epsilon_\mathrm{qDRIFT}=0.05$. Clearly, if we aim to simulate the Hubbard model for a longer time period, other optimization techniques must be considered in order to upgrade the current Bonsai mapping.

\section{BONSAI SIMULATED ANNEALING}
\label{app:dimostrazione}

In this section, we report the implementation details adopted to find the 
optimal Bonsai string assignment described in Section~\ref{Bonsai optimization}.
The algorithm is initialized from a naive assignment in which each Majorana 
string generated by the ternary-tree construction is associated with the 
fermionic site carrying the same index.

At each step, two sites $i$ and $j$ are selected on which we perform an 
exchange of the Majorana pairs $(m_{2i}, m_{2i+1})$ and $(m_{2j}, m_{2j+1})$.
Since each proposed swap modifies only the Hamiltonian terms that explicitly 
involve sites $i$ and $j$, the cost variation $\Delta\mathcal{C}$ can be 
computed via a local update restricted to $\mathcal{O}(n)$ terms, rather 
than recomputing the full cost function over all $\mathcal{O}(n^2)$ 
Hamiltonian terms. This reduces the total computational cost from 
$\mathcal{O}(n^2 \cdot N_{\mathrm{sweep}})$ to 
$\mathcal{O}(n \cdot N_{\mathrm{sweep}})$.

The new configuration is accepted with probability
\begin{equation}
  P = \begin{cases}
        1 & \text{if } \Delta\mathcal{C} < 0, \\
        \exp\!\left(-\beta(t)\,\Delta\mathcal{C}\right) & \text{if } 
        \Delta\mathcal{C} > 0,
      \end{cases}
\end{equation}
where
\begin{equation}
\beta(t) = A(t) \cdot c_1 \cdot c_2^{-t} \cdot \Theta(t - 1000)
\end{equation}
is the annealing schedule introduced in Section~\ref{Bonsai optimization}, 
and
\begin{equation}
    A(t) = 1 - c_3 \cdot \left|\sin\!\left(\frac{\pi t}{T}\right)\right|
\end{equation}
is a modulator function that periodically warms up the system every $T$ 
proposed configurations, allowing the algorithm to escape local minima 
during the early stages of the optimization, while progressively favoring 
lower-cost configurations as the search proceeds. Here $\Theta$ denotes 
the Heaviside step function. Although the method does not guarantee global 
optimality, performing $N_{\mathrm{sweep}} = 10^{5}$ iterations enables an 
extensive exploration of the Bonsai configuration space. For our 
optimization, we used $c_1 = 50$, $c_2 = 0.99995$, $c_3 = 0.4$, and 
$T = 2000$.

\begin{algorithm*}[!ht]
\caption{Simulated Annealing for Majorana String Assignment}
\label{alg:simulated_annealing}
\KwIn{Hamiltonian $H$, hardware graph $\mathcal{G}$, set of Majorana strings $\mathcal{S}$, cost function $\mathcal{C}$, annealing schedule $\beta$, number of sweeps $N_{\text{sweep}}$}
\KwOut{Optimized Hamiltonian $H_{\text{best}}$ and assignment list $L_{\text{best}}$ of Majorana string pairs to lattice sites}
\BlankLine
Initialize naive configuration\;
$L \leftarrow [1, \ldots, n_{\text{sites}}]$\;
$\mathcal{C}_{\text{best}} \leftarrow \mathcal{C}(H)$\;
\BlankLine
\For{$t = 0$ \KwTo $N_{\text{sweep}}$}{
    \For{$i = 1$ \KwTo $n_{\text{sites}}$}{
        Randomly sample $j \in [1, n_{\text{sites}}]$\;
        Construct $H'$ by exchanging Majorana pairs $[m_{2i}, m_{2i+1}]$ with $[m_{2j}, m_{2j+1}]$\;
        Construct $L'$ by swapping indices $i$ and $j$ in $L$\;
        Compute $\Delta\mathcal{C} \leftarrow \mathcal{C}(H') - \mathcal{C}(H)$\;
        \BlankLine
        \If{$\Delta\mathcal{C} \leq 0$ \textbf{or} $e^{-\beta(t)\,\Delta\mathcal{C}} > u \sim \mathcal{U}(0,1)$}{
            $H \leftarrow H'$\;
            $L \leftarrow L'$\;
            \If{$\mathcal{C}(H) < \mathcal{C}_{\text{best}}$}{
                $H_{\text{best}} \leftarrow H$\;
                $L_{\text{best}} \leftarrow L$\;
                $\mathcal{C}_{\text{best}} \leftarrow \mathcal{C}(H)$\;
            }
        }
    }
}
\BlankLine
\Return $H_{\text{best}}$, $L_{\text{best}}$\;
\end{algorithm*}

\section{DETAILED SIMULATION RESULTS OF THE HUBBARD MODEL}
\label{app:sim_results}

\begin{table*}[t]
\caption{%
  Classical exact simulation of the 2D Hubbard model on a $3\times 3$
  lattice with hopping $J=0.5$ and on-site interaction $U=2$ ($U/J=4$),
  starting from the N\'{e}el state.
  Local occupation numbers $\langle n_{i,\uparrow}\rangle$,
  $\langle n_{i,\downarrow}\rangle$, and local magnetization
  $\langle S^z_i\rangle = \langle n_{i,\uparrow}\rangle -
  \langle n_{i,\downarrow}\rangle$ are reported for each site $i$
  at four evolution times.
  Uncertainties are $\pm1\sigma$; exact integer values carry no
  uncertainty.
  The bottom block of each panel reports global observables;
  the  qDRIFT error bound $\Delta(\mathcal{O})_\mathrm{qDRIFT}$ associated to a given observable $\mathcal{O}$
  is given in the last row of the lower panel for each observable considered.
}
\label{tab:classical_3x3}
\centering
 
\begin{tabular*}{\textwidth}{@{\extracolsep{\fill}} ll ccc ccc @{}}
\toprule
 & &
  \multicolumn{3}{c}{$t_\mathrm{ev} = 0$} &
  \multicolumn{3}{c}{$t_\mathrm{ev} = 0.025, \; N = 79$} \\
\cmidrule(lr){3-5}\cmidrule(lr){6-8}
Site & Spin &
  $\langle n_\uparrow\rangle$ & $\langle n_\downarrow\rangle$ & $\langle S^z\rangle$ &
  $\langle n_\uparrow\rangle$ & $\langle n_\downarrow\rangle$ & $\langle S^z\rangle$ \\
\midrule
1 & $\uparrow$ &
  $1$ & $0$ & $+1$ &
  $0.9980 \pm 0.0014$ & $0.0020 \pm 0.0014$ & $0.996 \pm 0.003$ \\
2 & $\downarrow$ &
  $0$ & $1$ & $-1$ &
  $0.006 \pm 0.002$ & $0.996 \pm 0.002$ & $-0.990 \pm 0.004$ \\
3 & $\uparrow$ &
  $1$ & $0$ & $+1$ &
  $0.999 \pm 0.001$ & $0.009 \pm 0.003$ & $0.990 \pm 0.004$ \\
4 & $\downarrow$ &
  $0$ & $1$ & $-1$ &
  $0$ & $0.9970 \pm 0.0017$ & $-0.9970 \pm 0.0017$ \\
5 & $\uparrow$ &
  $1$ & $0$ & $+1$ &
  $0.9990 \pm 0.0010$ & $0$ & $0.9990 \pm 0.0010$ \\
6 & $\downarrow$ &
  $0$ & $1$ & $-1$ &
  $0.007 \pm 0.003$ & $0.986 \pm 0.004$ & $-0.979 \pm 0.006$ \\
7 & $\uparrow$ &
  $1$ & $0$ & $+1$ &
  $0.9980 \pm 0.0014$ & $0.005 \pm 0.002$ & $0.993 \pm 0.004$ \\
8 & $\downarrow$ &
  $0$ & $1$ & $-1$ &
  $0.0010 \pm 0.0010$ & $0.995 \pm 0.002$ & $-0.994 \pm 0.003$ \\
9 & $\uparrow$ &
  $1$ & $0$ & $+1$ &
  $0.9970 \pm 0.0017$ & $0.0020 \pm 0.0014$ & $0.995 \pm 0.003$ \\
\midrule
\multicolumn{2}{l}{$\langle N_\mathrm{tot}\rangle$} &
  \multicolumn{3}{c}{$9$} &
  \multicolumn{3}{c}{$ 9.00 \pm 0.03$} \\
\multicolumn{2}{l}{$\langle M_\mathrm{tot}\rangle$} &
  \multicolumn{3}{c}{$1$} &
  \multicolumn{3}{c}{$1.01 \pm 0.03 $} \\
\multicolumn{2}{l}{$\langle E\rangle$} &
  \multicolumn{3}{c}{$-4.5 \pm 0.6$} &
  \multicolumn{3}{c}{$-4.5 \pm 0.6 $} \\
\multicolumn{2}{l}{$\langle M_\mathrm{s}\rangle$} &
  \multicolumn{3}{c}{$ 1 $} &
  \multicolumn{3}{c}{$0.993 \pm 0.003$} \\
\bottomrule
\end{tabular*}

\begin{tabular*}{\textwidth}{@{\extracolsep{\fill}} ll ccc ccc @{}}
\toprule
 & &
  \multicolumn{3}{c}{$t_\mathrm{ev} = 0.05,\; N = 316$} &
  \multicolumn{3}{c}{$t_\mathrm{ev} = 0.125,\; N = 1969$} \\
\cmidrule(lr){3-5}\cmidrule(lr){6-8}
Site & Spin &
  $\langle n_\uparrow\rangle$ & $\langle n_\downarrow\rangle$ & $\langle S^z\rangle$ &
  $\langle n_\uparrow\rangle$ & $\langle n_\downarrow\rangle$ & $\langle S^z\rangle$ \\
\midrule
1 & $\uparrow$ &
  $0.9970 \pm 0.0017$ & $0.0020 \pm 0.0014$ & $0.995 \pm 0.003$ &
  $0.963 \pm 0.006$ & $0.0370 \pm 0.0060$ & $0.926 \pm 0.012$ \\
2 & $\downarrow$ &
  $0.990 \pm 0.003$ & $0.013 \pm 0.004$ & $0.977 \pm 0.007$ &
  $0.054 \pm 0.007$ & $0.972 \pm 0.005$ & $-0.918 \pm 0.012$ \\
3 & $\uparrow$ &
  $0.990 \pm 0.003$ & $0.013 \pm 0.004$ & $0.977 \pm 0.007$ &
   $0.964 \pm 0.006$ & $0.040 \pm 0.006$ & $0.924 \pm 0.012$ \\
4 & $\downarrow$ &
  $0.005 \pm 0.002$ & $0.994 \pm 0.002$ & $-0.989 \pm 0.005$ &
  $0.060 \pm 0.008$ & $0.950 \pm 0.007$ & $-0.890 \pm 0.014$ \\
5 & $\uparrow$ &
  $0.993 \pm 0.003$ & $0.007 \pm 0.003$ & $0.986 \pm 0.005$ &
 $0.928 \pm 0.008$  & $0.038 \pm 0.006$ & $0.890 \pm 0.014$ \\
6 & $\downarrow$ &
  $0.004 \pm 0.002$ & $0.984 \pm 0.004$ & $-0.980 \pm 0.006$ &
  $0.042 \pm 0.006$ & $0.930 \pm 0.008$ & $-0.888 \pm 0.014$ \\
7 & $\uparrow$ &
  $1$ & $0.009 \pm 0.003$ & $0.991 \pm 0.003$ &
  $0.970 \pm 0.005$ & $0.034 \pm 0.006$ &  $0.936 \pm 0.011$\\
8 & $\downarrow$ &
  $0.010 \pm 0.003$ & $0.995 \pm 0.002$ & $-0.985 \pm 0.005$ &
  $0.052 \pm 0.007$ & $0.972 \pm 0.005$ & $-0.920 \pm 0.012$ \\
9 & $\uparrow$ &
  $0.9970 \pm 0.0017$ & $0.013 \pm 0.004$ & $0.984 \pm 0.005$ &
  $0.964 \pm 0.006$ & $0.024 \pm 0.005$ & $0.940 \pm 0.011$ \\
\midrule
\multicolumn{2}{l}{$\langle N_\mathrm{tot}\rangle$} &
  \multicolumn{3}{c}{$9.01 \pm 0.04$} &
  \multicolumn{3}{c}{$ 8.99 \pm 0.11 $} \\
\multicolumn{2}{l}{$\langle M_\mathrm{tot}\rangle$} &
  \multicolumn{3}{c}{$0.99 \pm 0.04$} &
  \multicolumn{3}{c}{$1.00 \pm 0.11$} \\
\multicolumn{2}{l}{$\langle E\rangle$} &
  \multicolumn{3}{c}{$-4.5 \pm 0.6$} &
  \multicolumn{3}{c}{$-4.6 \pm 0.8$} \\
\multicolumn{2}{l}{$\langle M_\mathrm{s}\rangle$} &
  \multicolumn{3}{c}{$0.986 \pm 0.005$} &
  \multicolumn{3}{c}{$0.915 \pm 0.013$} \\
\midrule
\multicolumn{3}{l}{$ \Delta(\mathcal{O})_\mathrm{qDRIFT}$} &
  \multicolumn{1}{c}{$\langle n \rangle, \langle m \rangle, \langle M_\mathrm{s}\rangle  \sim 8 \cdot10^{-3}$} &
  \multicolumn{1}{c}{$\langle M_\mathrm{tot} \rangle \sim 7.2 \cdot 10^{-2}$} &
  \multicolumn{1}{c}{$\langle N_\mathrm{tot} \rangle \sim 1.44 \cdot 10^{-1}$} &
  \multicolumn{1}{c}{$\langle E \rangle \sim 2.52 \cdot10^{-1}$}\\
\bottomrule
\end{tabular*}
\end{table*}

\begin{figure*}[t]
    \centering
    \centering
    \includegraphics[width=2.0\columnwidth]{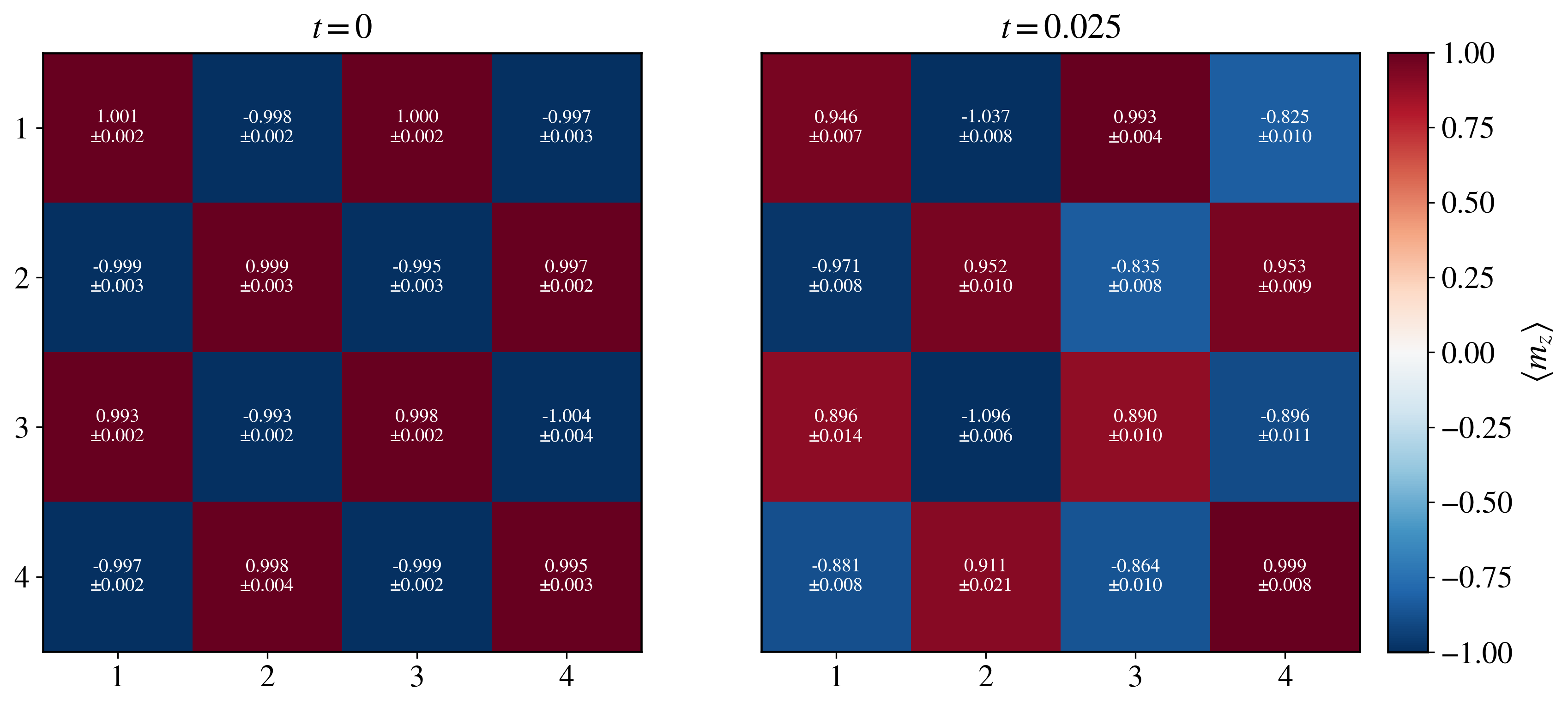}
    \includegraphics[width=2.0\columnwidth]{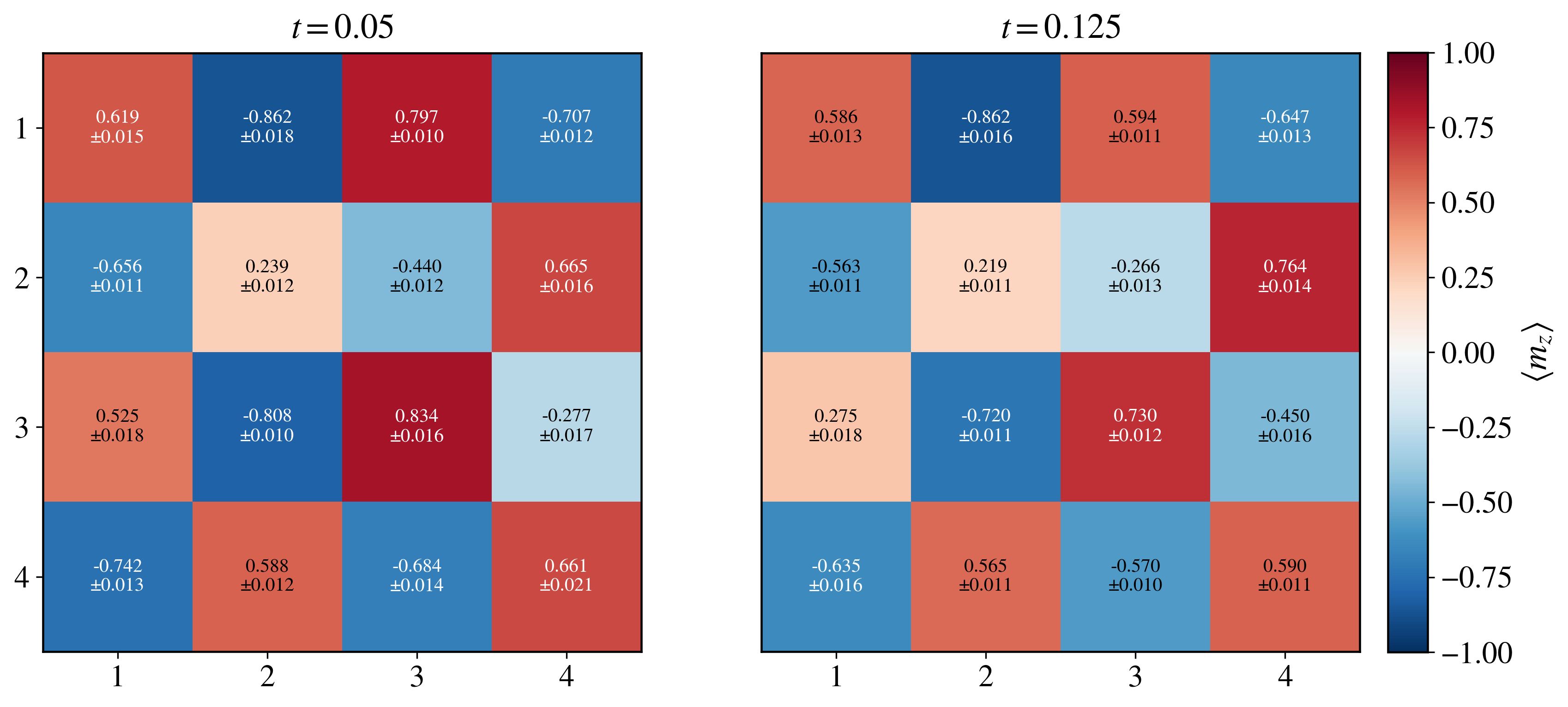}
    \caption{Local magnetization on a $4 \times 4$ squared lattice for $t=0$, $t=0.025$ (top figure), $t=0.05$ and $t=0.125$ (bottom figure). The initial Néel state is correctly prepared using the Bonsai mapping of \cite{Miller}, and evolve with time through the qDRIFT protocol. Despite the gate infidelity introduces noise in the circuit and the results quantitatively differs from the ideal simulation proposed in Table \ref{tab:classical_3x3} and the results obtained in \cite{alam2025programmabledigitalquantumsimulation}, the antiferromagnetic structure qualitatively remain preserved during the time evolution.}
    \label{fig:magnetizzazioni}
\end{figure*}

In this Appendix, we analyze additional information provided by the numerical simulation of the $3 \times 3$ lattice (ideal case) for the four times considered (in a Neél state initialization) in order to further verify the correctness of the algorithm designed in our work. Subsequently, we show the extrapolation results registered for the local magnetizations of the $4 \times 4$ to highlight how hardware noise affects the quality of the measurements as we time evolve the system.

Looking at Tab.~\ref{tab:classical_3x3}, for $t = 0$ the Néel state is initialized correctly within the Bonsai mapping, as confirmed 
by the occupation numbers. During the time evolution, virtual hopping 
corrections cause some fermions to hop onto already occupied sites. In 
particular, already at $t = 0.125$, the system begins to depart from its 
antiferromagnetic configuration, driven by the competition between the 
kinetic exchange term and the on-site repulsion, while keeping the total energy constant at $-4.5$, as expected from the conservation laws. 
Even though not reported here, the spatial 
correlation function exhibits the expected behavior: the alternating sign 
as a function of distance is generally preserved, and the value of the 
structure factor confirms that the system remains in an antiferromagnetic 
regime throughout the evolution.

Once the functioning of the code has been verified, it is possible to compare the local magnetization $\langle m_z(i) \rangle$ of the $4 \times 4$ lattice, shown in Fig.~\ref{fig:magnetizzazioni} as a function of site index and evolution time, with the values reported in Tab. \ref{tab:classical_3x3}. However, it is important to specify that a $3 \times 3$ lattice and a $4 \times 4$ lattice are generally different systems, as in the prior case not all the spins are paired, while in the latter the total magnetization is zero. Then, in principle, they have slightly different evolutions, even if they can be exploited for a qualitative analysis on the extrapolated results from the quantum hardware.

Each lattice site of the $4 \times 4$ system is represented as a checkerboard square, with red (blue) denoting positive (negative) magnetization.

At $t = 0$, the antiferromagnetic N\'{e}el order is clearly resolved across all 16 sites through the alternating color pattern, with measured values consistent with the ideal $|\langle m_z \rangle| = 1$ to within statistical uncertainty for 14 out of 16 sites. The residual deviations at sites 9 and 10 (counted from the top-left corner) are attributed to hardware noise during state preparation, and do not appreciably affect the global quality of the initial state: as shown in Fig.~\ref{fig:global_observables}, all global observables at $t = 0$ are statistically consistent with the ideal values within $1\sigma$.

As time evolves, hardware noise becomes increasingly dominant, particularly for the larger $6 \times 6$ lattice.

\section{ANALYSIS OF RICHARDSON EXTRAPOLATION}
\label{app:richardson}

In this appendix, we discuss the choice of the most suitable Richardson 
extrapolation scheme to mitigate the noise in our simulations.
We begin by summarizing the optimized Richardson protocol.

Let $\rho_{\lambda}$ denote the output of an imperfect quantum channel, 
where $\lambda$ is a dimensionless noise parameter quantifying the level 
of noise in the circuit. The expectation value of an observable 
$\mathcal{O}$ is given by
\begin{equation}
    E_{\lambda} = \Tr(\rho_{\lambda}\mathcal{O}). 
\end{equation}
Since the quantum channel is imperfect, the noise level $\lambda$ cannot 
be reduced below some finite value $\lambda_0 > 0$, and the zero-noise 
limit
\begin{equation}
    \lim_{\lambda \rightarrow 0} E(\lambda) = E^{*}
\end{equation}
cannot be accessed experimentally. However, it is always possible to 
artificially increase $\lambda$ and then extrapolate the expectation value 
of $\mathcal{O}$ to the zero-noise limit.

Theoretically, the Richardson extrapolation is ideal if the functional dependency of the observable on the noise parameter is not analytically known.
In our case of interest, we assume that during the  computing process, the qubits are not entirely affected by pure Markovian noise, but heavy-hexagon processors could also develop non-Markovian deviations during the execution of the circuit \cite{mark1,mark2,mark3}. If we take into account of those effects, it is hard to design an analytical model of the noise, and an optimized procedure similar to that proposed by \cite{PhysRevA.106.062436} is necessary.

In the following, we 
parameterize the noise level as $\lambda = \lambda_0 x$, with 
$x \in [1,\infty)$. The idea behind Richardson extrapolation is to sample 
$E_{\lambda_0}(x)$ at $n+1$ distinct points $x_0, \dots, x_n$ and to 
find the unique degree-$n$ polynomial that interpolates these samples. 
Evaluating this polynomial at $x = 0$ yields an estimate $\hat{R}_n$ of 
the noiseless expectation value $E^{*}$, which can be written as a linear 
combination of the samples:
\begin{equation}
    \hat{R}_n = \sum^n_{j=0} \gamma_j\, E_{\lambda_0}(x_j),
\end{equation}
where
\begin{equation}
    \gamma_j = \prod_{k \neq j}\frac{x_j}{x_k - x_j}
\end{equation}
are the Lagrange weights evaluated at $x = 0$.

The bias $B[\hat{R}_n] = \hat{R}_n - E^{*}$ of this estimator is given by
\begin{equation}
    B[\hat{R}_n] = (-1)^n E^{(n+1)}_{\lambda_0}(\xi) 
    \frac{C_n}{(n+1)!},
    \label{eq:bias}
\end{equation}
where $E^{(n+1)}_{\lambda_0}(x)$ denotes the $(n+1)$-th derivative of 
$E_{\lambda_0}(x)$ and $C_n = \prod^n_{j=0} x_j$.

In addition to the bias, a statistical error arises from the finite number 
of measurements. The variance of the estimator is
\begin{equation}
    \Var[\hat{R}_n] = \sum^n_{j=0} \gamma_j^2 \frac{\sigma^2}{n_j},
\end{equation}
where $n_j$ is the number of shots allocated to node $x_j$ and $\sigma$ 
is the measurement variance, assumed uniform across nodes 
($\sigma_j = \sigma$) for simplicity. For a fixed total shot 
budget $\sum^n_{j=0} n_j = n_{\mathrm{tot}}$, the 
variance is minimized by allocating~\cite{hoel1964optimal,
cai2023practicalframeworkquantumerror}
\begin{equation}
    n_j = n_{\mathrm{tot}}\frac{|\gamma_j|}{\sum^{n}_{k=0}|\gamma_k|}.
\end{equation}
Defining $\Lambda = \sum^n_{j=0} |\gamma_j|$ and 
$n_{\mathrm{eff}} = n_{\mathrm{tot}}/\Lambda^2$, the minimized variance 
reduces to
\begin{equation}
    \Var[\hat{R}_n] = \frac{\sigma^2}{n_{\mathrm{tot}}}\Lambda^2 
    = \frac{\sigma^2}{n_{\mathrm{eff}}}.
\end{equation}

Furthermore, \cite{PhysRevA.106.062436} shows that the 
\textit{tilted Chebyshev nodes}
\begin{equation}
    x_j = 1 + \frac{\sin^2\!\left(\frac{j}{n+1}\frac{\pi}{2}\right)}
    {\sin^2\!\left(\frac{1}{n+1}\frac{\pi}{2}\right)}(x_1-1), 
    \quad j = 0,\dots,n,
\end{equation}
minimize the product $C_n$ in the bias Eq.~(\ref{eq:bias}) for a 
fixed $\Lambda$.

In principle, increasing $\Lambda$ reduces the bias at the expense of a 
larger shot budget $n_{\mathrm{tot}}$ required to keep the variance 
constant. However, large values of $\Lambda$ also exacerbate the classical 
Runge phenomenon: the nodes cluster close together, producing large 
Lagrange weights $\gamma_j$, so that even small fluctuations in a measured 
observable can cause numerical instability. Conversely, choosing $\Lambda$ 
too small results in noisy extrapolations for high-degree polynomials \cite{PhysRevA.106.062436}.

\bibliography{references}

@article{Jiang,
   title={Optimal fermion-to-qubit mapping via ternary trees with applications to reduced quantum states learning},
   volume={4},
   ISSN={2521-327X},
   url={http://dx.doi.org/10.22331/q-2020-06-04-276},
   DOI={10.22331/q-2020-06-04-276},
   journal={Quantum},
   publisher={Verein zur Forderung des Open Access Publizierens in den Quantenwissenschaften},
   author={Jiang, Zhang and Kalev, Amir and Mruczkiewicz, Wojciech and Neven, Hartmut},
   year={2020},
   month=jun, pages={276} }

@article{Feynman,
  author    = {R. Feynman},
  title     = {Simulating physics with computers},
  journal   = {International Journal of Theoretical Physics},
  volume    = {21},
  pages     = {467},
  year      = {1982}
}

@misc{AtomComputing_2023,
  title     = {Atom Computing Announces \textgreater 1{,}000 Qubit Quantum Computer},
  author    = {{Atom Computing}},
  year      = {2023},
  url       = {https://arstechnica.com/science/2023/10/atom-computing-is-the-first-to-announce-a-1000-qubit-quantum-computer/},
  note      = {Ars Technica report},
}

@misc{IBM_Condor_2023,
  title     = {IBM Condor Quantum Processor Surpasses 1{,}000 Qubits},
  author    = {{IBM Research}},
  year      = {2023},
  url       = {https://www.ibm.com/quantum/blog/quantum-roadmap-2033},
  note      = {IBM Quantum Summit 2023 announcement},
}

@article{Miller,
   title={Bonsai Algorithm: Grow Your Own Fermion-to-Qubit Mappings},
   volume={4},
   ISSN={2691-3399},
   url={http://dx.doi.org/10.1103/PRXQuantum.4.030314},
   DOI={10.1103/prxquantum.4.030314},
   number={3},
   journal={PRX Quantum},
   publisher={American Physical Society (APS)},
   author={Miller, Aaron and Zimborás, Zoltán and Knecht, Stefan and Maniscalco, Sabrina and García-Pérez, Guillermo},
   year={2023},
   month=aug }

@article{JW,
  author    = {P. Jordan and E. Wigner},
  title     = {\"Uber das Paulische \"Aquivalenzverbot},
  journal   = {Z. Phys.},
  volume    = {47},
  pages     = {631},
  year      = {1928}
}

@article{Bravyi,
   title={Fermionic Quantum Computation},
   volume={298},
   ISSN={0003-4916},
   url={http://dx.doi.org/10.1006/aphy.2002.6254},
   DOI={10.1006/aphy.2002.6254},
   number={1},
   journal={Annals of Physics},
   publisher={Elsevier BV},
   author={Bravyi, Sergey B. and Kitaev, Alexei Yu.},
   year={2002},
   month=may, pages={210–226} }

@misc{bravyi2017taperingqubitssimulatefermionic,
      title={Tapering off qubits to simulate fermionic Hamiltonians}, 
      author={Sergey Bravyi and Jay M. Gambetta and Antonio Mezzacapo and Kristan Temme},
      year={2017},
      eprint={1701.08213},
      archivePrefix={arXiv},
      primaryClass={quant-ph},
      url={https://arxiv.org/abs/1701.08213}, 
}

@misc{fischer2019symmetryconfigurationmappingrepresenting,
      title={Symmetry Configuration Mapping for Representing Quantum Systems on Quantum Computers}, 
      author={Sean A. Fischer and Daniel Gunlycke},
      year={2019},
      eprint={1907.01493},
      archivePrefix={arXiv},
      primaryClass={quant-ph},
      url={https://arxiv.org/abs/1907.01493}, 
}

@article{Steudtner_2018,
   title={Fermion-to-qubit mappings with varying resource requirements for quantum simulation},
   volume={20},
   ISSN={1367-2630},
   url={http://dx.doi.org/10.1088/1367-2630/aac54f},
   DOI={10.1088/1367-2630/aac54f},
   number={6},
   journal={New Journal of Physics},
   publisher={IOP Publishing},
   author={Steudtner, Mark and Wehner, Stephanie},
   year={2018},
   month=jun, pages={063010} }

@misc{IBM_HeavyHex_2021,
  author    = {P. Nation and H. Paik and A. Cross and Z. Nazario},
  title     = {The IBM Quantum Heavy Hex Lattice},
  year      = {2021},
  month     = jan,
  url       = {https://www.ibm.com/quantum/blog/heavy-hex-lattice},
  note      = {Accessed: 2025-06-20}
}

@article{Yu_2025,
   title={Clifford Circuit-Based Heuristic Optimization of Fermion-To-Qubit Mappings},
   volume={21},
   ISSN={1549-9626},
   url={http://dx.doi.org/10.1021/acs.jctc.5c00794},
   DOI={10.1021/acs.jctc.5c00794},
   number={19},
   journal={Journal of Chemical Theory and Computation},
   publisher={American Chemical Society (ACS)},
   author={Yu, Jeffery and Liu, Yuan and Sugiura, Sho and Van Voorhis, Troy and Zeytinoğlu, Sina},
   year={2025},
   month=sep, pages={9430–9443} }

@misc{david2025tightererrorboundsqdrift,
      title={Tighter Error Bounds for the qDRIFT Algorithm}, 
      author={I. J. David and I. Sinayskiy and F. Petruccione},
      year={2025},
      eprint={2506.17199},
      archivePrefix={arXiv},
      primaryClass={quant-ph},
      url={https://arxiv.org/abs/2506.17199}, 
}

@article{Campbell_2019,
   title={Random Compiler for Fast Hamiltonian Simulation},
   volume={123},
   ISSN={1079-7114},
   url={http://dx.doi.org/10.1103/PhysRevLett.123.070503},
   DOI={10.1103/physrevlett.123.070503},
   number={7},
   journal={Physical Review Letters},
   publisher={American Physical Society (APS)},
   author={Campbell, Earl},
   year={2019},
   month=aug }

@article{Seeley_2012,
   title={The Bravyi-Kitaev transformation for quantum computation of electronic structure},
   volume={137},
   ISSN={1089-7690},
   url={http://dx.doi.org/10.1063/1.4768229},
   DOI={10.1063/1.4768229},
   number={22},
   journal={The Journal of Chemical Physics},
   publisher={AIP Publishing},
   author={Seeley, Jacob T. and Richard, Martin J. and Love, Peter J.},
   year={2012},
   month=dec }

@article{Vlasov_2022,
   title={Clifford Algebras, Spin Groups and Qubit Trees},
   volume={11},
   ISSN={1314-7374},
   url={http://dx.doi.org/10.12743/quanta.v11i1.199},
   DOI={10.12743/quanta.v11i1.199},
   journal={Quanta},
   publisher={Quanta},
   author={Vlasov, Alexander Yurievich},
   year={2022},
   month=dec, pages={97–114} }

@article{suzuki,
  author    = {M. Suzuki},
  title     = {Fractal decomposition of exponential operators with applications to many-body theories and Monte Carlo simulations},
  journal   = {Phys. Lett. A},
  volume    = {146},
  pages     = {319},
  year      = {1990}
}

@article{suzuki2,
  author    = {M. Suzuki},
  title     = {General theory of fractal path integrals with applications to many‐body theories and statistical physics},
  journal   = {J. Math. Phys.},
  volume    = {32},
  pages     = {400},
  year      = {1991}
}

@article{Berry_2006,
   title={Efficient Quantum Algorithms for Simulating Sparse Hamiltonians},
   volume={270},
   ISSN={1432-0916},
   url={http://dx.doi.org/10.1007/s00220-006-0150-x},
   DOI={10.1007/s00220-006-0150-x},
   number={2},
   journal={Communications in Mathematical Physics},
   publisher={Springer Science and Business Media LLC},
   author={Berry, Dominic W. and Ahokas, Graeme and Cleve, Richard and Sanders, Barry C.},
   year={2006},
   month=dec, pages={359–371} }

@article{annealing,
  author    = {S. Kirkpatrick and C. Gelatt and M. Vecchi},
  title     = {Optimization by simulated annealing},
  journal   = {Science},
  volume    = {220},
  pages     = {671},
  year      = {1983}
}

@article{annealing1,
  author    = {P. van Laarhoven and E. Aarts},
  title     = {Simulated annealing: Theory and Applications},
  journal   = {Springer Netherlands:Dordrecht},
  pages     = {7},
  year      = {1987}
}

@article{annealing2,
  author    = {D. Bertsimas and J. Tsitsiklis},
  title     = {Simulated annealing.},
  journal   = {Statistical Science},
  volume    = {8},
  pages     = {10},
  year      = {1993}
}

@incollection{annealing3,
  author    = {Nikolaev, A. and Jacobson, S.},
  title     = {Simulated Annealing},
  booktitle = {Handbook of Metaheuristics},
  editor    = {Gendreau, M. and Potvin, J.-Y.},
  publisher = {Springer},
  address   = {New York},
  year      = {2010},
  pages     = {1}
}

@article{Takahashi1980Steiner,
  author  = {Takahashi, H. and Matsuyama, A.},
  title   = {An approximate solution for the Steiner problem in graphs},
  journal = {Mathematica Japonicae},
  volume  = {24},
  pages   = {573--577},
  year    = {1980},
  month   = jan,
  url     = {https://pascal-francis.inist.fr/vibad/index.php?action=getRecordDetail&idt=PASCAL8030328467},
}

@INPROCEEDINGS{shrub,
  author={Grewal, G. and Wilson, T. and Xu, M. and Banerji, D.},
  booktitle={17th International Conference on VLSI Design. Proceedings.}, 
  title={Shrubbery: a new algorithm for quickly growing high-quality Steiner trees}, 
  year={2004},
  volume={},
  number={},
  pages={855-862},
  doi={10.1109/ICVD.2004.1261038}}

@book{Steger2002Steiner,
  author    = {Hans J{\"u}rgen Pr{\"o}mel and Angelika Steger},
  title     = {The Steiner Tree Problem: A Tour through Graphs, Algorithms, and Complexity},
  year      = {2002},
  publisher = {Vieweg},
  address   = {Braunschweig}
}

@article{PhysRevA.106.062436,
  title = {Optimization of Richardson extrapolation for quantum error mitigation},
  author = {Krebsbach, Michael and Trauzettel, Bj\"orn and Calzona, Alessio},
  journal = {Phys. Rev. A},
  volume = {106},
  issue = {6},
  pages = {062436},
  numpages = {10},
  year = {2022},
  month = {Dec},
  publisher = {American Physical Society},
  doi = {10.1103/PhysRevA.106.062436},
  url = {https://link.aps.org/doi/10.1103/PhysRevA.106.062436}
}

@misc{cai2023practicalframeworkquantumerror,
      title={A Practical Framework for Quantum Error Mitigation}, 
      author={Zhenyu Cai},
      year={2023},
      eprint={2110.05389},
      archivePrefix={arXiv},
      primaryClass={quant-ph},
      url={https://arxiv.org/abs/2110.05389}, 
}

@article{hoel1964optimal,
  author          = {Hoel, Paul G. and Levine, A.},
  title           = {Optimal Spacing and Weighting in Polynomial Prediction},
  journal    = {The Annals of Mathematical Statistics},
  volume          = {35},
  number          = {4},
  pages           = {1553--1560},
  year            = {1964},
  doi             = {10.1214/aoms/1177700382},
  url             = {https://doi.org},
  publisher       = {Institute of Mathematical Statistics}
}

@misc{mark1,
      title={Probing non-Markovian qubit noise and modeling Post Markovian Master Equation}, 
      author={Chun-Tse Li and Jingming Tan and Vasil Gucev and Daniel Lidar},
      year={2025},
      eprint={2510.12894},
      archivePrefix={arXiv},
      primaryClass={quant-ph},
      url={https://arxiv.org/abs/2510.12894}, 
}

@article{mark2,
  title = {Unifying Non-Markovian Characterization with an Efficient and Self-Consistent Framework},
  author = {White, G. A. L. and Jurcevic, P. and Hill, C. D. and Modi, K.},
  journal = {Phys. Rev. X},
  volume = {15},
  issue = {2},
  pages = {021047},
  numpages = {43},
  year = {2025},
  month = {May},
  publisher = {American Physical Society},
  doi = {10.1103/PhysRevX.15.021047},
  url = {https://link.aps.org/doi/10.1103/PhysRevX.15.021047}
}

@article{mark3,
  title = {Characterizing non-Markovian off-resonant errors in quantum gates},
  author = {Wei, Ken Xuan and Pritchett, Emily and Zajac, David M. and McKay, David C. and Merkel, Seth},
  journal = {Phys. Rev. Appl.},
  volume = {21},
  issue = {2},
  pages = {024018},
  numpages = {19},
  year = {2024},
  month = {Feb},
  publisher = {American Physical Society},
  doi = {10.1103/PhysRevApplied.21.024018},
  url = {https://link.aps.org/doi/10.1103/PhysRevApplied.21.024018}
}

@misc{alam2025programmabledigitalquantumsimulation,
      title={Programmable digital quantum simulation of 2D Fermi-Hubbard dynamics using 72 superconducting qubits}, 
      author={Faisal Alam and Jan Lukas Bosse and Ieva Čepaitė and Adrian Chapman and Laura Clinton and Marcos Crichigno and Elizabeth Crosson and Toby Cubitt and Charles Derby and Oliver Dowinton and Paul K. Faehrmann and Steve Flammia and Brian Flynn and Filippo Maria Gambetta and Raúl García-Patrón and Max Hunter-Gordon and Glenn Jones and Abhishek Khedkar and Joel Klassen and Michael Kreshchuk and Edward Harry McMullan and Lana Mineh and Ashley Montanaro and Caterina Mora and John J. L. Morton and Dhrumil Patel and Pete Rolph and Raul A. Santos and James R. Seddon and Evan Sheridan and Wilfrid Somogyi and Marika Svensson and Niam Vaishnav and Sabrina Yue Wang and Gethin Wright},
      year={2025},
      eprint={2510.26845},
      archivePrefix={arXiv},
      primaryClass={quant-ph},
      url={https://arxiv.org/abs/2510.26845}, 
}

@article{SciPostPhysCodeb.41,
	title = {Tensor network Python (TeNPy) version 1},
	pages = {41},
	author = {Hauschild, Johannes and Unfried, Jakob and Anand, Sajant and Andrews, Bartholomew and Bintz, Marcus and Borla, Umberto and Divic, Stefan and Drescher, Markus and Geiger, Jan and Hefel, Martin and Hémery, Kévin and Kadow, Wilhelm and Kemp, Jack and Kirchner, Nico and Liu, Vincent S. and Moller, Gunnar and Parker, Daniel and Rader, Michael and Romen, Anton and Scalet, Samuel and Schoonderwoerd, Leon and Schulz, Maximilian and Soejima, Tomohiro and Thoma, Philipp and Wu, Yantao and Zechmann, Philip and Zweng, Ludwig and Mong, Roger and Zaletel, Michael P. and Pollmann, Frank},
	journal = {SciPost Phys. Codebases},
	year = {2024},
	publisher = {SciPost},
	doi = {10.21468/SciPostPhysCodeb.41},
	url = {https://scipost.org/10.21468/SciPostPhysCodeb.41}
}

@article{Dijkstra1959,
  author   = {E. W. Dijkstra},
  title    = {A note on two problems in connexion with graphs},
  journal  = {Numerische Mathematik},
  year     = {1959},
  volume   = {1},
  number   = {1},
  pages    = {269},
  doi      = {10.1007/BF01386390},
  url      = {https://doi.org/10.1007/BF01386390},
}

@article{georgescu2014quantum,
  title={Quantum simulation},
  author={Georgescu, Iulia M and Ashhab, Sahel and Nori, Franco},
  journal={Reviews of Modern Physics},
  volume={86},
  number={1},
  pages={153--185},
  year={2014},
  publisher={APS}
}

@article{low2017optimal,
  title={Optimal Hamiltonian simulation by quantum signal processing},
  author={Low, Guang Hao and Chuang, Isaac L},
  journal={Physical review letters},
  volume={118},
  number={1},
  pages={010501},
  year={2017},
  publisher={APS}
}

@article{motlagh2024generalized,
  title={Generalized quantum signal processing},
  author={Motlagh, Danial and Wiebe, Nathan},
  journal={PRX Quantum},
  volume={5},
  number={2},
  pages={020368},
  year={2024},
  publisher={APS}
}

@article{zimboras2025myths,
  title={Myths around quantum computation before full fault tolerance: What no-go theorems rule out and what they don't},
  author={Zimbor{\'a}s, Zolt{\'a}n and Koczor, B{\'a}lint and Holmes, Zo{\"e} and Borrelli, Elsi-Mari and Gily{\'e}n, Andr{\'a}s and Huang, Hsin-Yuan and Cai, Zhenyu and Ac{\'\i}n, Antonio and Aolita, Leandro and Banchi, Leonardo and others},
  journal={arXiv preprint arXiv:2501.05694},
  year={2025}
}

\end{document}